\documentclass[ACM-Reference-Format,numbers,10.5pt,compsoc]{BD}

\usepackage{graphicx}
\usepackage{footmisc}
\usepackage{subfigure}
\usepackage{url}
\usepackage{multirow}
\usepackage{amsmath,amsthm}
\usepackage{amssymb,amsfonts}
\usepackage{booktabs}
\usepackage{color}
\usepackage{ccaption}
\usepackage{booktabs}
\usepackage{float}
\usepackage{fancyhdr}
\usepackage{caption}
\usepackage{xcolor,stfloats}
\usepackage{comment}
\graphicspath{{figures/}}
\usepackage{cuted}  
\usepackage{captionhack}
\usepackage{epstopdf}
\usepackage{pdfpages}
\usepackage{lscape}
\usepackage{threeparttable}
\usepackage{arydshln}
\usepackage{adjustbox}
\usepackage{longtable}
\usepackage{makecell}
\usepackage{subfigure}
\usepackage[figuresright]{rotating}
\usepackage{cite}

\headevenname{\zihao{-5}{\textbf{\emph{Big Data Mining and Analytics, January}}} 2018, 1(1): 000-000}%
\headoddname{{\sf Anxin Yang et al.:}\quad {\textbf{\emph{A Survey of Reasoning for Substitution Relationships: Defnitions, Methods, and Directions}}}}%

\newtheoremstyle{mystyle}{0pt}{0pt}{\normalfont}{1em}{\bf}{}{1em}{}
\theoremstyle{mystyle}
\renewcommand\figurename{Fig.~}

\newcommand{\nop}[1]{}

\makeatletter
\renewcommand{\@biblabel}[1]{[#1]\hfill}
\makeatother

\begin{document}

\thispagestyle{empty}

\hyphenpenalty=50000

\makeatletter
\newcommand\mysmall{\@setfontsize\mysmall{7}{9.5}}

\newenvironment{tablehere}
  {\def\@captype{table}}
  {}
\newenvironment{figurehere}
  {\def\@captype{figure}}
  {}

\thispagestyle{plain}%
\thispagestyle{empty}%

\let\temp\footnote
\renewcommand \footnote[1]{\temp{\zihao{-5}#1}}
{}
\vspace*{-40pt}
\noindent{\zihao{5-}\textbf{\scalebox{0.95}[1.0]{\makebox[5.9cm][s]
{BIG\hfill  DATA \hfill MINING \hfill AND \hfill ANALYTICS}}}}

\vskip .2mm
{\zihao{5-}
\textbf{
\hspace{-5mm}
\scalebox{1}[1.0]{\makebox[5.6cm][s]{%
I\hspace{0.70pt}S\hspace{0.70pt}S\hspace{0.70pt}N\hspace{0.70pt}{\color{white}%
2\hspace{-2pt}2\hspace{0.70pt}}2\hspace{0.70pt}0\hspace{0.70pt}9\hspace{0.70pt}6\hspace{0.70pt}-\hspace{0.70pt%
}0\hspace{0.70pt}6\hspace{0.70pt}5\hspace{0.00pt}4\hspace{0.70pt}\hspace{0.70pt%
}\hspace{0.70pt}{\color{white}l\hspace{0.70pt}l\hspace{0.70pt}}0\hspace{0.70pt}%
?\hspace{0.70pt}/\hspace{0.70pt}?\hspace{0.70pt}?\hspace{0.70pt}{\color{white}%
l\hspace{0.70pt}l\hspace{0.70pt}}p\hspace{0.70pt}p\hspace{0.70pt}?\hspace{0.70pt}?\hspace{0.70pt}?%
--\hspace{ 0.70pt}?\hspace{0.70pt}?\hspace{0.70pt}?}}}

\vskip .2mm\noindent
{\zihao{5-}\textbf{\scalebox{1}[1.0]{\makebox[5.6cm][s]{%
V\hspace{0.4pt}o\hspace{0.4pt}l\hspace{0.4pt}u\hspace{0.4pt}m\hspace{0.4pt}%
e\hspace{0.4em}1\hspace{0.4pt},\hspace{0.8em}N\hspace{0.4pt}u\hspace{0.4pt}%
m\hspace{0.4pt}b\hspace{0.4pt}e\hspace{0.4pt}r\hspace{0.4em}1,\hspace{0.8em}%
J\hspace{0.4pt}a\hspace{0.4pt}n\hspace{0.4pt}u\hspace{0.4pt}a\hspace{0.4pt}%
\hspace{0.4pt}r\hspace{0.4pt}y\hspace{0.4em}2\hspace{0.4pt}0\hspace{0.4pt}1\hspace{0.4pt}8}}}}

\vskip .2mm\noindent
{\zihao{5-}\textbf{\scalebox{1}[1.0]{\makebox[5.6cm][s]{%
\color{white}{V\hfill o\hfill l\hfill u\hfill m\hfill%
e\hspace{0.356em}1,\hspace{0.356em}N\hfill u\hfill%
m\hfill b\hfill e\hfill r\hspace{0.356em}1,\hspace{0.356em}%
S\hfill e\hfill p\hfill t\hfill e\hfill%
m\hfill b\hfill e\hfil lr\hspace{0.356em}2\hfill0\hfill1\hfill8}}}}}\\

\begin{strip}
{\center
{\zihao{3}\textbf{
A Survey of Reasoning for Substitution Relationships: Definitions, Methods, and Directions
}}
\vskip 9mm}

{\center {\sf \zihao{5}
Anxin Yang, Zhijuan Du, and Tao Sun $^*$
}
\vskip 5mm}
%

\centering{
\begin{tabular}{p{160mm}}

{\zihao{-5}
\linespread{1.6667} %
\noindent
\bf{Abstract:} {\sf
Substitute relationships play a vital role in people’s daily lives, covering various fields. This study focuses on the understanding and prediction of substitute relationships of products in different domains, comprehensively analyzing the application of machine learning algorithms, natural language processing, and other technologies. By comparing the model methods in various directions such as the definition of substitutes in different domains, representation and learning of substitute relationships, and substitute reasoning, it provides a methodological basis for in-depth exploration of substitute relationships. Through continuous research and innovation, we can further enhance the personalization and accuracy of substitute recommendation systems, thereby promoting the development and application of this field.}
}

\end{tabular}
}
\vskip 6mm

\vskip -3mm
\zihao{6}\end{strip}

\thispagestyle{plain}%
\thispagestyle{empty}%
\makeatother
\pagestyle{tstheadings}

\begin{figure}[b]
\vskip -6mm
\begin{tabular}{p{44mm}}
\toprule\\
\end{tabular}
\vskip -4.5mm
\noindent
\setlength{\tabcolsep}{1pt}
\begin{tabular}{p{1.5mm}p{79.5mm}}

$\bullet$& Anxin Yang with School of Computer Science, Inner Mongolia University, Hohhot City, Inner Mongolia, 010021, China. E-mail: 32209097@mail.imu.edu.cn \\
$\bullet$& Zhijuan Du with School of Computer Science, Inner Mongolia University, Hohhot City, Inner Mongolia, 010021, China. E-mail: nmg-duzhijuan@163.com \\
$\bullet$& Tao Sun with School of Computer Science, Inner Mongolia University, Hohhot City, Inner Mongolia, 010021, China. E-mail: cssunt@imu.edu.cn \\
 \\
$\sf{*}$&
To whom correspondence should be addressed. \\
          &          Manuscript received: year-month-day;
          accepted: year-month-day

\end{tabular}
\end{figure}\zihao{5}

\vbox{}
\vskip 1mm
\noindent

\section{INTRODUCTION}

In today's diverse market environment, consumers have the option to choose different products or services as alternatives to their original choices. The concept of substitution has become increasingly important, as it offers selectivity and flexibility in meeting one's needs.
\begin{itemize}
	\item[$\bullet$]\textbf{Case 1: }A user wants to purchase a specific brand and model of a smartphone, but discovers that it is sold out on the local e-commerce website. In such a situation, the user needs to find a substitute with similar features and performance, begins comparing other brands and models of smartphones. They compare various specifications, functions, design, and user reviews to select the most suitable phone for their needs. They may also consider other factors such as after-sales service, brand reputation, and reliability. Ultimately, they choose to purchase a smartphone from another brand as a substitute. This case highlights the behavior of consumers in the retail industry when a particular product fails to meet their needs, as they seek alternatives with similar features or characteristics. It demonstrates the flexibility and decision-making abilities of consumers in ensuring they obtain the products or services they require.
	
	\item[$\bullet$]\textbf{Case 2: }Due to food allergies, religious beliefs, health requirements, or personal preferences, people may need to choose alternative ingredients based on specific requirements. If a recipe calls for a particular type of fish to make a dish, but this fish is not available in the local market, the person may choose another delicious fish with a similar texture as a substitute to ensure that the dish's taste and flavor are not greatly affected. Alternatively, if someone plans to make a dessert but runs out of the required chocolate, they may choose to substitute it with cocoa powder and butter to achieve a similar taste and texture.
\end{itemize}

Consumers may harbor varied preferences and needs across different shopping scenarios, allowing them to choose among disparate products offering similar functionalities. For instance, when purchasing a television, consumers can make their selection based on personal preferences for brand, price, display technology, and other factors. Furthermore, substitution relationships can also provide alternative options when a particular product is temporarily out of stock or discontinued. For instance, when a certain product is temporarily unavailable or phased out, consumers can opt for other products with similar functionalities or features. In terms of food, people may need to choose alternative ingredients based on specific requirements due to food allergies, religious beliefs, health needs, or personal preferences. For example, for vegetarians, they may opt for alternatives such as legumes, mushrooms, or plant-based meat products as substitutes for animal meat.

\subsection{Characterizations and Applications of Substitution}

``Substitute" typically refers to the nature of offering items with similar functionalities or characteristics when a particular item fails to meet the user's needs. These items, which possess substitutability, are referred to as substitutes. Explaining the process of substitution can aid in understanding why these substitutes are recommended and assist people in making better decisions.

\textbf{In the retail industry: }As demonstrated by the aforementioned case 1, substitutes refer to alternative goods that possess similar functionality or features in meeting the users' needs. When a specific product fails to fulfill the users' requirements, individuals tend to seek alternative products that offer comparable functionality or characteristics as replacements.

\textbf{In the field of food:} In case 2, specific application scenarios are mentioned. Substitutes are primarily defined based on the nutritional composition, taste, and texture of the food. They may exhibit similar flavors (such as sweet or sour), comparable textures (like crisp or fluffy), belong to the same food category (for instance, substituting different varieties of potatoes), or be used in similar recipe contexts (for example, utilizing bacon or chicken as the primary protein in a sandwich).

\textbf{In other fields:} Similarity and applicability are factors considered in determining substitutes. For instance, in fields like pharmaceuticals and business investment, the search for substitutes often involves seeking alternatives that exhibit a high degree of similarity.

\begin{table*}[htbp] 
	\centering
	\resizebox{\linewidth}{!}{ 
		\begin{tabular}{@{}lcccl@{}} 
			\toprule 
			Domain & Data Source for Substitutes & Criteria for Extracting Substitutes & Pros & Cons \\ 
			\midrule 
			Retail industry & Price elasticity & Increase in demand for item B due to price rise of item A & Precise identification of substitutes & High data requirement \\ 
			Retail industry & Co-view, view-but-purchase-another &\makecell{ Users’ browsing data with “co-view” and \\ “view-but-purchase-another” patterns }& \makecell{ Simple judgment\\ easy data acquisition }& Label noise\\ 
			The field of food & User reviews & Statements in reviews containing “substitute” information & Simple judgment & Inconsistent standards \\
			The field of food & Recipes & – & Strong relevance to results & Low universality \\
			Other fields & Similarity & – & – & – \\  
			\bottomrule 
			
		\end{tabular} 
	}
	\caption{Pros and cons of the approach to extracting substitutes.} 
		\label{1_tab}
\end{table*}

The characterization methods for substitutes are shown in Table \ref{1_tab}. Overall, similarity is an important criterion for determining substitutes, as objects with substitutability have a high degree of similarity to each other.

In the retail industry, using price elasticity to determine substitutes is an economic method used to measure the impact of price changes on consumer demand and to identify other products that could potentially substitute for the original product. Price elasticity analysis can accurately determine substitutes for the original product in the market. However, conducting accurate price elasticity analysis requires a large amount of market data and consumer behavior data, as well as a substantial amount of consumption data. On the other hand, considering co-browsing behavior in the retail industry can provide a simple method of assessment. By analyzing users’ co-browsing behavior during online shopping, it is possible to infer their level of interest in a particular product.However, the limitation of this method lies in the relatively easy acquisition of data, but the issue of label noise can lead to inaccuracies in predictions\cite{Ye_2023_Transformer-Based}. In the food industry, using user reviews as an alternative data source is a simple operation, from which words such as "substitute a for b ” or “replace b with a”\cite{Dai_2022_Decomposing}.However, user reviews vary in terms of standards, making it challenging to have a benchmark for comparing results from different methods. This makes it difficult to make accurate judgments based on such data.By choosing substitute ingredients based on the recipe, high-quality substitutes with similar cooking effects and textures can be obtained. However, there is a strong correlation between dishes and ingredients, and different dishes cannot be universally applicable.In other fields, when data is lacking, it is common to search for highly similar items as substitutes through the analysis of their characteristic features. This process often involves manual review to ensure the accuracy of the data results.

Each method has its unique advantages and disadvantages, requiring a balance and consideration of the specific problem and the characteristics of the data to choose the most appropriate solution.

\subsection{Research Value and Related Surveys }
The search for substitutes can be achieved through substitute reasoning to meet personalized needs and better adaptation. Due to differences in taste and regional culture, people have diverse perceptions of food substitutes, making it difficult to find a satisfying solution that applies to everyone. Achieving recipe personalization through ingredient substitution may help people meet their dietary needs and preferences, avoid potential allergens, and simplify the process of culinary exploration. In order to address the issue of ingredient substitution, in the recommendation of food substitutes, we can use small-scale data learning to search for substitute foods with similar characteristics. For product substitution, we can use machine learning and automation technology to identify which products can substitute for the target product, or recommend related substitutes based on customer feedback. These methods and technologies not only help consumers to be more flexible and convenient in purchasing goods and cooking processes, but also provide more accurate recommendation services and brand marketing for businesses.

The substitution relationship provides consumers and decision-makers with the flexibility and options to make product choices in different scenarios, enabling them to make the best decisions based on their individual needs and goals. Study \cite{Kwark__Spillover} conducted research on the influence of online product reviews among different brands of substitute products across a wide range of product categories on consumer purchasing decisions. The study revealed that, for substitute products, the majority of positive reviews have the potential to enhance product evaluation while potentially diminishing the appeal of other substitute products, thus shifting preferences towards the focal substitute product.Research on consumer behavior indicates that\cite{Zhang__Complements} when the recommended product's price exceeds that of the focal product, it raises consumers' psychological expectation of the price. During the product browsing stage, consumers tend to prefer comparing multiple substitutes, making it crucial to identify the complementary and substitutive relationships between products. Unlike previous research that primarily focused on the relationship between online reviews and sales figures, a study \cite{Yin_2021_How} has demonstrated the impact of the richness of online reviews (i.e., including videos or subsequent comments) on sales. The findings reveal that utilitarian products exhibit a stronger influence from online reviews compared to hedonic products, and negative reviews have a more pronounced effect on product perception than positive ones.

Starting from \cite{McAuley_2015_Inferring}, the accuracy of product recommendations has been improved by analyzing substitute relationships and complementary relationships. Substitute relationships refer to the extent to which two products can replace each other. Complementary relationships refer to the ability of two products to complement or be used together. For example, cookies and milk have a complementary relationship because milk can be consumed with cookies to enhance the user’s eating experience. Recommendation systems can utilize complementary relationships to provide corresponding recommendations and increase user purchase intention \cite{10027703,10.1145/3604915.3608864,Kang2018CompleteTL}.

The relationship reasoning technology of products can infer the correlation and connection between products based on the user's historical behavior and preferences. By analyzing user's purchase history, click behavior, ratings, and other data, it is possible to build a network of relationships between products. This network can describe relationships such as similarity, substitution, and complementarity between products.Based on these relationships, recommendation systems can utilize product relationship reasoning to help users browse through a large collection of products and find items that are relevant to their interests. Researchers have been working tirelessly to propose new models to improve the performance of product recommendation systems \cite{Zhang2020GraphbasedRO,Ding2020ImprovingIR,Yan2021LkeRecTL}. In recommendation systems, personalization and data sparsity have always been important aspects of concern, and some recently proposed models attempt to provide better solutions to these problems. Substitute relationships and complementary relationships are vital concepts that play a crucial role in improving recommendation accuracy and personalization by mining the similarity and connections between products. This, in turn, enhances the precision of recommendations and user satisfaction.

Visual similarity and substitute relationships are both used to describe the relationships between things, and they are used to compare the degree of similarity between two objects or concepts. Both visual similarity and substitute relationships have a certain degree of subjectivity, as people's judgments of similarity and substitute relationships may vary due to individual experiences, cultural backgrounds, and cognitive preferences. However, visual similarity primarily focuses on comparing the degree of similarity between things based on visual features such as appearance, structure, shape, color, etc., i.e., the similarity based on external features. On the other hand, substitute relationships primarily focus on whether one object or concept can replace another in a specific context, i.e., the relationship based on substitutability. In the field of product recommendation, research on substitute relationships is often based on the visual features of the products, such as the style of product images\cite{McAuley_2015_Image-Based}.Subsequent research often focuses on analyzing images in a specific domain, such as the fashion and clothing industry \cite{Cheng2020FashionMC,Gu2020FashionAA}. This is because in the fashion and clothing industry, the visual features of products play a crucial role in consumers' purchasing decisions.Therefore, researchers typically apply image analysis to this specific domain to uncover substitute and complementary relationships between products\cite{Lin2019FashionOC,Pachoulakis2012AUGMENTEDRP,Hwangbo2018RecommendationSD}. This survey primarily focuses on the inference of general product relationships in recommendation systems, with less emphasis on image analysis in the fashion domain.

\subsection{Framework for Organizing Substitute Work }
A general framework for substitute reasoning includes data preprocessing, feature representation, substitute relation learning and inference models, model training and optimization, as well as model evaluation.

During the\textbf{ data preprocessing stage}, the data is cleansed, standardized, and transformed to ensure its quality and consistency.\textbf{Feature representation}, a crucial component of the substitute reasoning framework, involves converting the elements and attributes of the data into an effective representation. This may involve traditional feature engineering methods such as statistical features or manually defined attributes, or it can leverage deep learning techniques for automatic feature extraction. The quality and effectiveness of feature representation are vital for subsequent learning and inference processes. \textbf{Substitute relation learning and inference models} form the core part of the framework, enabling the discovery and inference of new substitute relationships, thereby further understanding the hidden information and structure within the data. Based on the data processed through feature representation, learning and inference models suitable for the substitute reasoning task can be designed and constructed. These models can be implemented through different machine learning algorithms, deep neural networks, or graph neural networks, among other techniques.During the \textbf{model training and optimization stage}, it involves selecting appropriate objective functions, tuning model parameters, and choosing suitable optimization methods to ensure the model can better adapt to the training data and possesses good generalization capabilities. Finally, \textbf{ model evaluation} utilizes appropriate datasets and evaluation metrics to measure the performance of the model, such as accuracy, recall, F1 score, and so on.

This survey aims to study and explore the application of methods for inferring substitute relationships, the specific structure is shown in the following diagram.

\begin{figure*}
	\centering 
	\includegraphics[width=0.85\textwidth]{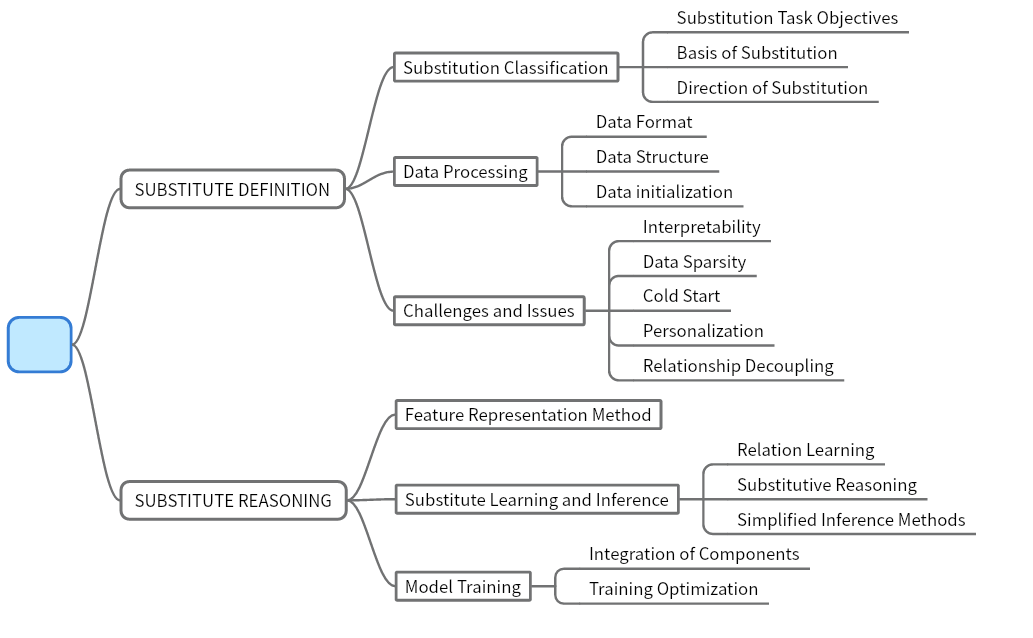}
	\caption{Structural Diagram}
\end{figure*}

\begin{itemize}
	\item[$\bullet$] Chapter 2 will focus on introducing the classification methods, data processing methods, and main problems addressed in the inference of substitute relationships. The classification of substitutes will be based on task objectives, underlying factors, and directions of substitution. Data processing methods are crucial in substitute relationship inference and involve preprocessing of data and data structures, among others. Additionally, this chapter will also provide a detailed discussion on the main problems addressed by substitute relationship inference, such as interpretability issues, data sparsity, cold start, and relationship decoupling.
	\item[$\bullet$] Chapter 3, the exploration delves deeper into the methods for feature representation, learning, reasoning, and model training associated with substitute relationships. Feature representation methods play a crucial role in effectively capturing the essential elements and attributes of causal relationships. The subsequent learning and reasoning methods aim to harness these features to uncover new substitute relationships, while model training methods, whether supervised or unsupervised, are essential in refining the inference process.
	\item[$\bullet$] Chapter 4 will focus on the commonly used datasets and evaluation standards for substitute relationship inference. Appropriate datasets and accurate evaluation standards are crucial for the study of inference methods. This chapter will introduce some commonly used datasets and discuss evaluation metrics for assessing the effectiveness of inference, such as accuracy, recall, and others.
	\item[$\bullet$]Finally,Chapter 5 reveals our future challenges , and Chapter 6 presents our conclusion. 
\end{itemize}

Through comprehensive analysis and discussion of substitute relationship inference methods, this survey aims to provide a comprehensive framework for researchers in substitute relationship inference. The goal is to better apply and promote its development.

\section{SUBSTITUTE DEFINITION}
``Substitute" refers to the existence of items, services, or concepts in a specific context that can be mutually replaced. Substitutes often occur based on observation and research into the direction and basis of substitution. For example, in the food industry, we can analyze the attributes and nutritional components of ingredients to determine which ones can be substituted for each other. In the realm of commodities, we can compare product features and functions to find products that can be substituted for each other.Establishing feature initialization forms and embedding spaces helps in analyzing and predicting substitution patterns and directions. It enables us to gain a better understanding and analysis of the relationships and characteristics of substitutions.

\begin{table*}[hb]
	\centering
	\resizebox{\linewidth}{!}{
		\begin{tabular}{cccccccccclcccccc} 
			\\
			\toprule 
			\multirow{3}{*}{Model}                                                      & \multicolumn{6}{c}{Definition of Substitutes}                                                                                                                                                & \multicolumn{3}{c}{}                                                & \multicolumn{7}{c}{Problem Solving}                                                                                                                                                    \\
			& \multicolumn{4}{c}{Substitute Task Objectives}                                                                                                                  & \multicolumn{2}{c}{Substitute Directions}                     & \multicolumn{3}{c}{Data Form and Structure}                                         & \multicolumn{5}{c}{Common Problems}                                                                                        & \multicolumn{2}{c}{Approaches}                                     \\
			& \multicolumn{1}{l}{\rotatebox{90}{Food Substitutes}} & \multicolumn{1}{l}{\rotatebox{90}{Product Substitutes}} & \multicolumn{1}{l}{Other Substitutes} & \multicolumn{1}{l}{\rotatebox{90}{Modeling Complementarity}}                                & \rotatebox{90}{One-way substitutes}                 &\rotatebox{90}{ Two-way substitutes}                  & \rotatebox{90}{Text data }                 & \rotatebox{90}{Image data}                 & \rotatebox{90}{Structured}             & \rotatebox{90}{Interpretability} & \multicolumn{1}{l}{\rotatebox{90}{Data sparsity}} & \multicolumn{1}{l}{\rotatebox{90}{Multilingual}} & \multicolumn{1}{l}{\rotatebox{90}{Cold start}} & \multicolumn{1}{l}{\rotatebox{90}{Personalization}} & \multicolumn{1}{l}{\rotatebox{90}{Decoupling multiple relations}} & \multicolumn{1}{l}{\rotatebox{90}{Path-based}} \\\midrule
			GISMo\cite{Fatemi_2023_Learning}                                                            & \checkmark                    &                       &                      &                                                           & \checkmark                    &                       & \checkmark                     & \checkmark                    &                      &   &                      &                       &                      & \checkmark                    &                       &                        \\\cite{Ye_2023_Transformer-Based}                                           &                        & \checkmark                    &                       &                                                             &                       &  \checkmark                     & \checkmark                    &                        &                        &   &                       & \checkmark                     &                        &                        &                       &                        \\
			\cite{Li__Food}                                                                & \checkmark                    &                       &                      &                                                           & \checkmark                    &                       & \checkmark                     &                      & KG                   &   & \checkmark                    &                       & \checkmark                    & \checkmark                    & \multicolumn{1}{l}{} & \multicolumn{1}{l}{} \\
			\cite{Lawrynowicz_2022_Food}                                                                & \checkmark                    &                       &                      &                                                           &                      &                       & \checkmark                     &                      & KG                   &   &                      &                       &                      &                      &                      &                      \\
			SCG-SPRe\cite{Zhang_2022_Learning}                                                         &                      & \checkmark                     &                      & \checkmark                                                         &                      & \checkmark                     & \checkmark                     &                      & Graph                     &   &                      &                       & \checkmark                    &                      & \checkmark                    &                      \\
			DHGAN\cite{Zhou_2022_Decoupled}                                                            &                      & \checkmark                     &                      & \checkmark                                                         &                      & \checkmark                     & \checkmark                     &                      & Graph                     & \checkmark & \checkmark                    &                       &                      &                      & \checkmark                    & \checkmark                    \\
			M-HetSage\cite{Jian_2022_Multi-task}                                                        &                      & \checkmark                     &                      &                                                           &                      & \checkmark                     & \checkmark                     & \checkmark                    & Graph                     &   &                      &                       &                      &                      &                      &                      \\
			SPGCN\cite{Dai_2022_Decomposing}                                                            & \multicolumn{1}{l}{} &                       & Corporate Investment                 & \checkmark                                                         &                      & \checkmark                     & \checkmark                     &                      & Graph                     &   &                      &                       &                      &                      & \checkmark                    &                      \\
			KAPR\cite{Yang_2022_Inferring}                                                             &                      & \checkmark                     &                      & \checkmark                                                         &                      & \checkmark                     & \checkmark                     &                      & KG                   & \checkmark & \checkmark                    &                       &                      &                      &                      & \checkmark                    \\
			Food2Vec/BERT\cite{Pellegrini_2021_Exploiting} & \checkmark                    &                       &                      &                                                           &                      &                       & \checkmark                     & \checkmark                    &                      &   &                      &                       &                      &                      &                      &                      \\
			DIISH\cite{Shirai_2021_Identifying}                                                            & \checkmark                    &                       &                      &                                                           &                      &                       & \checkmark                     &                      &                      &   &                      &                       &                      &                      &                      &                      \\
			A2CF\cite{Chen_2020_Try}                                                             &                      & \checkmark                     &                      &                                                           &                      & \checkmark                     & \checkmark                     &                      &                      & \checkmark &                      &                       &                      & \checkmark                    &                      &                      \\
			DecGCN\cite{Liu_2020_Decoupled}                                                           &                      & \checkmark                     &                      & \checkmark                                                         &                      & \checkmark                     & \checkmark                     &                      &                      &   &                      &                       &                      &                      & \checkmark                    &                      \\
			Product2Vec\cite{Chen_2020_Studying}                                                      &                      & \checkmark                     &                      & \checkmark                                                         &                      & \checkmark                     & \checkmark                     &                      &                      &   &                      &                       &                      &                      &                      &                      \\
			RRN\cite{Zhang_2019_Identifying}                                                              &                      & \checkmark                     &                      & \checkmark                                                         &                      & \checkmark                     & \checkmark                     &                      &                      &   &                      &                       &                      &                      &                      &                      \\
			KGDDS\cite{Shen_2019_KGDDS}                                                            & \multicolumn{1}{l}{} &                       & Medicine                   &                                                           &                      & \checkmark                     & \checkmark                     &                      & KG                   &   &                      &                       &                      &                      &                      &                      \\
			SPEM\cite{Zhang_2019_Inferring}                                                             &                      & \checkmark                     &                      &                                                           &                      & \checkmark                     & \checkmark                     &                      & Graph                    &   &                      &                       &                      &                      &                      &                      \\
			LVA\cite{Rakesh_2019_Linked}          &                      & \checkmark                     &                      &\checkmark&                      & \checkmark                     & \checkmark                     &                      &                      &   &                      &                       & \checkmark                    & \checkmark                    &                      &                      \\
			PMSC\cite{Wang_2018_Path-constrained}                                                             &                      & \checkmark                     &                      & \checkmark                                                         &          \checkmark            &                          & \checkmark                     &                      &                      &   &                      &                       & \checkmark                    &                      & \checkmark                    & \checkmark                    \\
			SHOPPER\cite{Ruiz_2019_SHOPPER}                                                          &                      & \checkmark                     &                      & \checkmark                                                         &                      & \checkmark                     & \checkmark                     &                      &                      &   & \checkmark                    &                       &                      & \checkmark                    &                      &                      \\
			Sceptre\cite{McAuley_2015_Inferring}                                                          &                      & \checkmark                     &                      & \checkmark& \checkmark                    &                       & \checkmark                     &                      & Graph                    &   &                      &                       & \checkmark                    &                      & \checkmark                    & \checkmark                    \\
			\cite{McAuley_2015_Image-Based}                                                                &                      & \checkmark                     &                      & \checkmark                                                         &                      &                       &                       & \checkmark                    &                      &   &                      &                       & \checkmark                    & \checkmark                    &                      &                      \\
			\cite{Boscarino_2014_Automatic}                                                                & \checkmark                    &                       &                      &                                                           &                      &                       & \checkmark                     &                      &                      &   &                      &                       &                      &                      &                      &\\ EMRIGCN\cite{Chen_2023_Enhanced}                                                                &                      & \checkmark                      &                      & \checkmark                                                          &                      & \checkmark                     & \checkmark                     &                      &  Graph                    &   &                      &                       & \checkmark                     &                      &                      &\\IRGNN\cite{Liu_2022_Item}   &  & \checkmark &  & \checkmark & \checkmark &   & \checkmark &  & Graph &  & \checkmark &  &   &  &  & \checkmark\\                     
			\bottomrule          
		\end{tabular}
}
	\caption{Definition and problems of substitute reasoning}
\end{table*}

\subsection{Substitution Classification}

\subsubsection{Substitution Task Objectives}
When the required items are unavailable or missing, people may need to look for alternative products to meet their needs, and the requirements for substitutions can vary.

For example, in the substitution of ingredients, some people may be allergic to certain ingredients and need to find substitutes \cite{Pellegrini_2021_Exploiting,Shirai_2021_Identifying,Boscarino_2014_Automatic}; Because different ingredients can have varying flavors and textures, substituting specified ingredients in a given recipe 
require some modifications to the recipe and cooking method as well\cite{Fatemi_2023_Learning,Li__Food}, it is important to consider additional factors such as nutritional value and cooking characteristics.

Research\cite{Ye_2023_Transformer-Based,Jian_2022_Multi-task,Chen_2020_Try} focuses on reasoning about substitution relationships between products. However, sometimes the task of relationship inference requires exploring multiple relationships. For example, the substitution relationship between products is often modeled together with complementary relationships.In research \cite{Chen_2023_Enhanced,Yang_2022_Inferring,Rakesh_2019_Linked,McAuley_2015_Inferring}, substitution relationships and complementary relationships are modeled separately, without considering the interaction between different relationships. However, some studies suggest that there is an interaction between substitution and complementary relationships\cite{Liu_2022_Item,Zhang_2022_Learning,Zhou_2022_Decoupled,Yang_2022_Inferring,Liu_2020_Decoupled,Chen_2020_Studying,Wang_2018_Path-constrained},capture the dependencies between relationships can provide a better understanding of the implicit relationships between products. The figure \ref{3-1-_img} shows the coupling patterns of product relationships:
\begin{figure}[htp]
	\centering
	\subfigure[]{
		\includegraphics[scale=0.24]{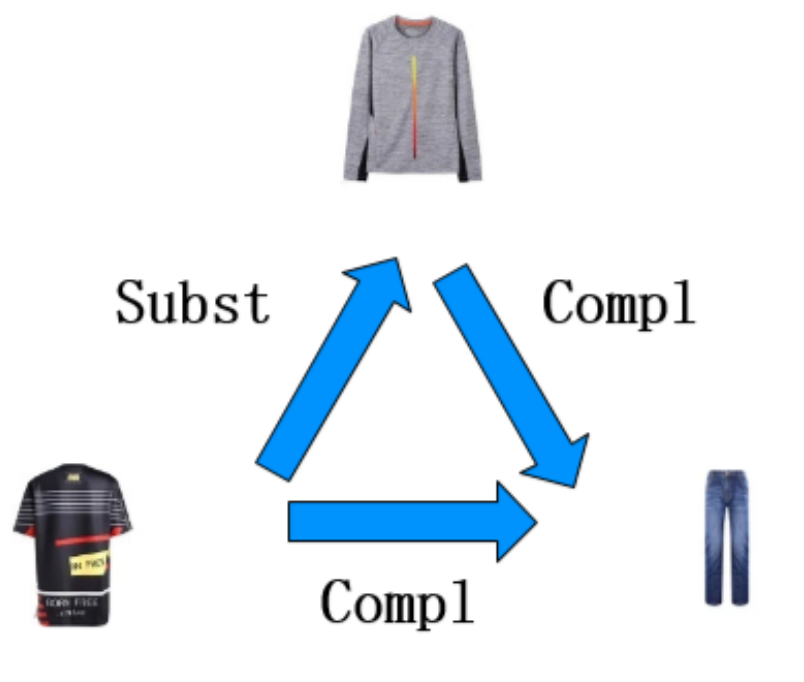}}
	\subfigure[]{
		\includegraphics[scale=0.24]{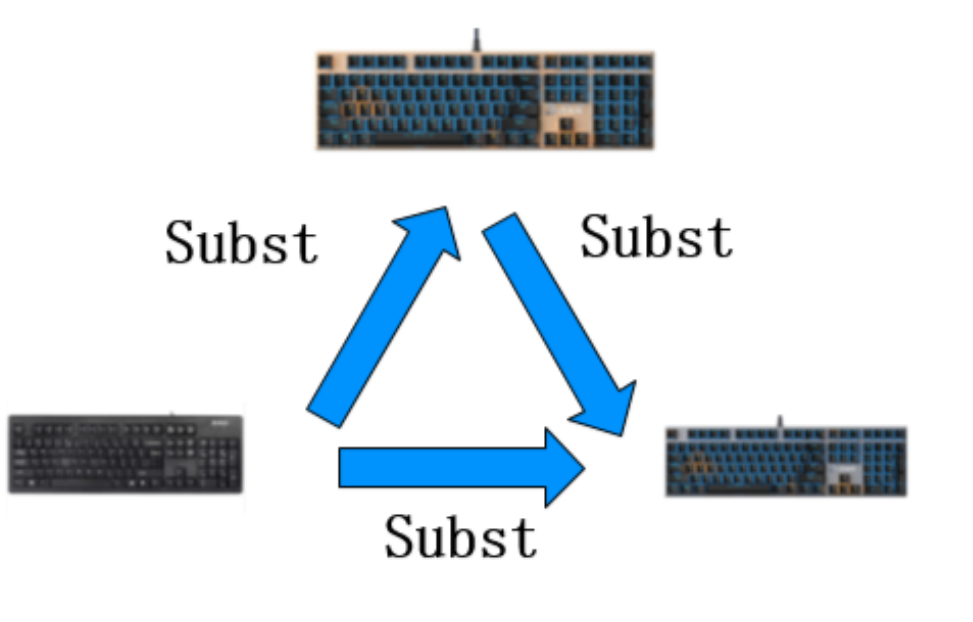}}
	\subfigure[]{
		\includegraphics[scale=0.24]{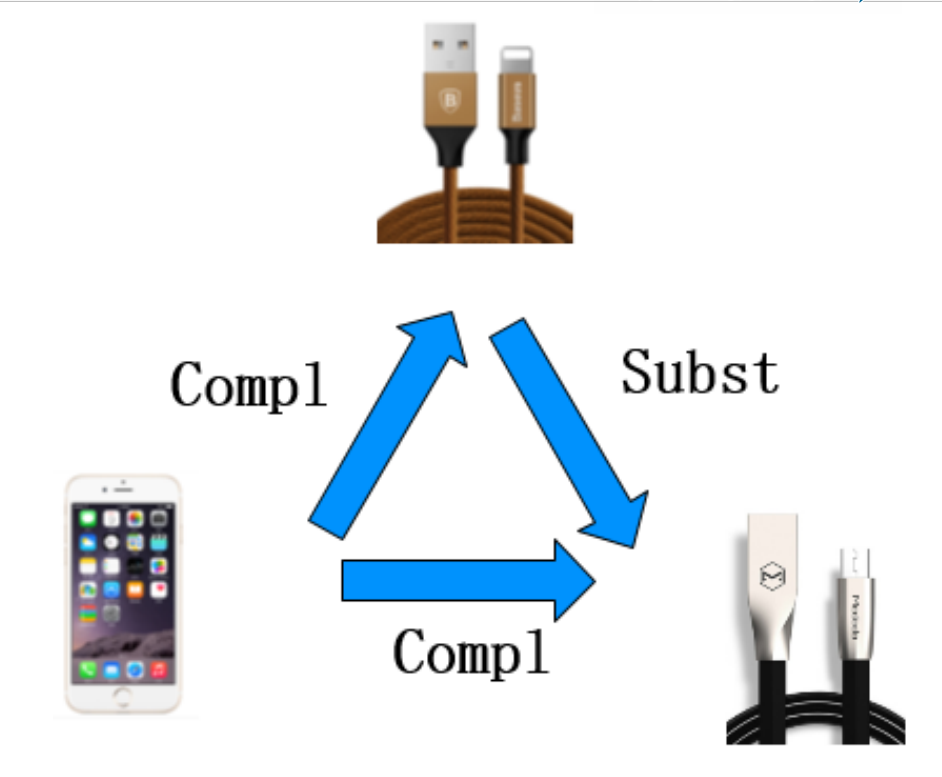}}
	\subfigure[]{
		\includegraphics[scale=0.24]{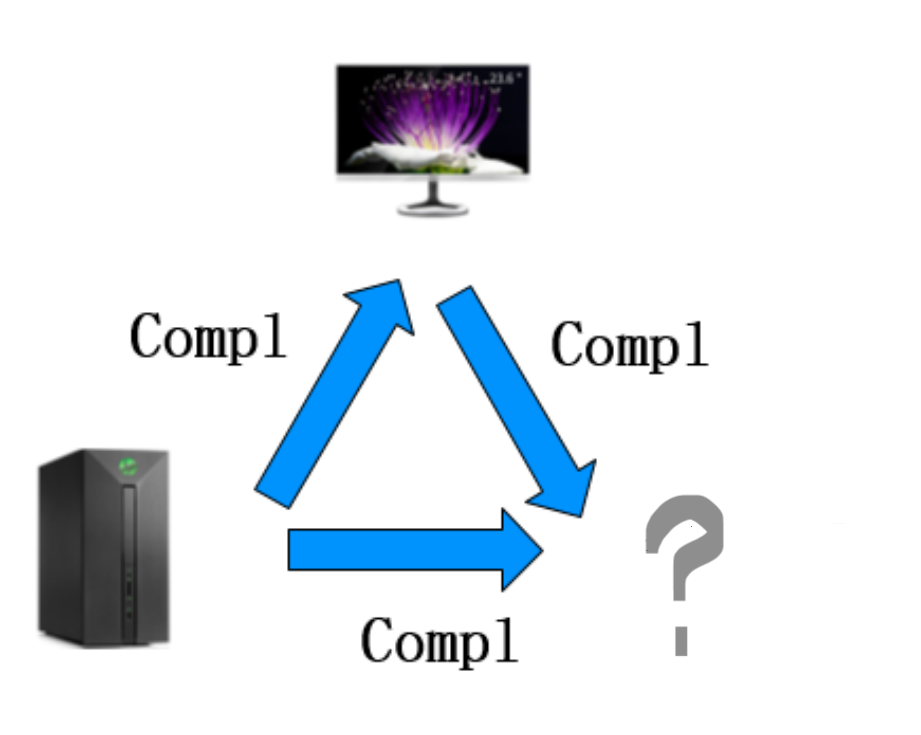}}
	\caption{Coupling patterns of relationships}
	\label{3-1-_img}
\end{figure}

\subsubsection{Basis of Substitution}
The basis for substitution refers to the selection and replacement of items, services, or concepts based on different conditions in a particular context. The basis for substitution can vary in different situations.

The concept \cite{Boscarino_2014_Automatic} of using rule-based deduction for ingredient substitution was first introduced, and in the study, ingredients were transformed into topic distributions, with substitutions selected based on the density of the distribution. In research \cite{Shirai_2021_Identifying,Pellegrini_2021_Exploiting,Li__Food,Lawrynowicz_2022_Food,Fatemi_2023_Learning}believes that the substitutability and similarity of ingredients are related, and suggests using the semantic aspect of the ingredients as the basis for substitution. The research \cite{Li__Food,Fatemi_2023_Learning} utilized the semantic aspect of ingredients while also considering the correlation of ingredients in different recipe contexts, determining substitutions based on specific scenario conditions and requirements; taking into account the similarity between different recipes and the co-occurrence of ingredients in recipes\cite{Shirai_2021_Identifying}, the obtained information is integrated as a basis for substitution reasoning; in addition, \cite{Fatemi_2023_Learning,Pellegrini_2021_Exploiting} incorporated ingredient images as auxiliary data, combining the semantic information from textual data with the visual features from images. By analyzing and processing the ingredient images, they were able to improve the accuracy of substitution choices.

In order to explore the substitution relationships between products, the study \cite{McAuley_2015_Image-Based} attempted to simulate the human perception of the visual characteristics of objects by analyzing the style of images for substitution reasoning.In addition to product similarity, the study \cite{McAuley_2015_Inferring} suggests that co-browsed products are substitutable. Researches \cite{Wang_2018_Path-constrained, Rakesh_2019_Linked, Zhou_2022_Decoupled, Liu_2020_Decoupled, Chen_2023_Enhanced, Liu_2022_Item} use existing product substitutions and complementary connections as the basis for relational inference, inferring their substitution relationships by learning the associations between products. Apart from co-browsing, other user behaviors also provide implicit feedback on the relationships between products. \cite{McAuley_2015_Inferring, Zhang_2019_Identifying} combine product reviews to predict relationships between products. Additionally, \cite{Zhang_2019_Identifying} also considers non-textual features of comments in their analysis; \cite{Zhang_2022_Learning} distinguishes relationships between products based on the temporal patterns of user behavior sequences; \cite{Jian_2022_Multi-task} explores product substitution relationships through user search logs and product stock-out logs. There is mutual influence between products, and \cite{Ruiz_2019_SHOPPER,Chen_2020_Studying} analyze user shopping baskets and orders to infer the substitutability of products based on their co-occurrence relationships.

And \cite{Dai_2022_Decomposing} discusses the selection of investment in enterprises, stating that the interdependencies between companies determine their substitutability. On the other hand, \cite{Shen_2019_KGDDS} defines substitution from the perspective of the effects and side effects of drugs.

\subsubsection{Direction of Substitution}
In practical applications, substitution often occurs in a bidirectional manner, however, not all substitution behaviors are bidirectional, and in some cases, substitution can only occur in a unidirectional manner.

For example, in some recipes, a specific ingredient is necessary to achieve the expected taste and texture. \cite{Fatemi_2023_Learning,Li__Food} discusses ingredient substitution and recipes, aiming to replace rare, allergenic, expensive, or unhealthy ingredients with their common, non-allergenic, cheaper, or healthier counterparts.

In e-commerce, the substitution relationship of goods is assumed to be bidirectional by default, but the specific situation still depends on the usage context of the goods. When a consumer selects a product, an e-commerce platform may display other similar products as alternative options, but the specific substitution relationship still depends on personalized factors such as consumer needs and individual preferences. \cite{McAuley_2015_Inferring,Liu_2022_Item} argue that there is a directionality in the relationships between products. \cite{McAuley_2015_Inferring} models the product relationship graph as a directed graph, while \cite{Liu_2022_Item} makes directional predictions using non-commutative outer product operations. The specific substitution relationship depends on the product itself, user needs, and the specific application context.

\subsection{Data Processing}

\subsubsection{Data Format and Structure}
Choosing appropriate data format and structure can enhance the representation efficiency and access speed of data, thereby optimizing program performance.

\textbf{Text Data:} Text data comes from various sources, such as ingredient names in the food domain\cite{Lawrynowicz_2022_Food}, recipes\cite{Boscarino_2014_Automatic}, and cooking steps\cite{Fatemi_2023_Learning}. Recipes serve as the contextual environment for ingredients and can help understand the correlation between ingredient names, providing more accurate recommendations and suggestions suitable for the current scenario. However, recipes are subject to cultural differences, and the lack of universality in ingredient substitutions. Cooking steps can ensure the quality and repeatability of recipes, but constructing a cooking model is not easy.

Common data sources used in product substitution include: product reviews\cite{Zhang_2019_Identifying}, product titles\cite{Chen_2020_Try}, product descriptions\cite{Yang_2022_Inferring}, user behavior\cite{Jian_2022_Multi-task}, purchase history\cite{Ruiz_2019_SHOPPER}, and tags\cite{Fatemi_2023_Learning}.Product titles provide basic product information, typically in a more general manner, while product descriptions provide more detailed information. On the other hand, if the features provided by product descriptions are too detailed, it may result in overfitting of the model, making it difficult to generalize to new data. Understanding user preferences and providing personalized product choices can be done by analyzing reviews and user browsing/ purchasing behavior to find alternative products and discover cross-selling opportunities. However, user behavior data often follows a long-tail distribution, where popular or common products have a large number of reviews, while reviews for some unique or rare products are scarce. Therefore, to comprehensively model user behavior and product features, it is necessary to integrate data from multiple perspectives, such as user behavior data and product information.

\textbf{Image data :}\cite{McAuley_2015_Image-Based} is used for modeling fashion style substitutions with other products. Some studies \cite{Fatemi_2023_Learning, Pellegrini_2021_Exploiting, Jian_2022_Multi-task} combine text embeddings with image embeddings to accurately understand and infer data through multimodal learning and information extraction.

Choosing the appropriate data structure is crucial for efficient data representation and processing. Compared to unstructured textual data, graphs composed of nodes and edges can visually reflect the relationships between products \cite{McAuley_2015_Inferring, Zhang_2019_Inferring}. Graphs provide a clear view of the complex relationships and network structures among products, enabling us to understand dependencies and interdependencies between products.Graph algorithms can be used for analysis and processing \cite{Dai_2022_Decomposing, Jian_2022_Multi-task, Zhou_2022_Decoupled}, enabling the discovery and utilization of patterns of association among products. Knowledge graphs, as a type of heterogeneous graph, offer a richer and more flexible modeling approach based on entities and relationships. By integrating data from different sources and types into a unified knowledge graph \cite{Shen_2019_KGDDS, Yang_2022_Inferring, Li__Food}, a more comprehensive understanding can be achieved.

\subsubsection{Data initialization}

For the process of feature initialization, researchers can use different methods and models to extract and represent the features of the data, in support of subsequent tasks and analyses.

In the realm of textual data, the LDA model can be utilized for topic modeling, extracting latent thematic features and performing low-dimensional embeddings on text, as well as embedding representations of product reviews \cite{McAuley_2015_Inferring, Zhang_2019_Identifying}. However, due to LDA's unsuitability for modeling short texts, the Word2Vec model acquires word vector representations by learning the distribution of words in context, capturing the semantic information of words through generating word vectors. Some studies \cite{Wang_2018_Path-constrained, Yang_2022_Inferring} have chosen to initialize data using variants of Word2Vec \cite{mikolov2013efficient}, which better reflect the semantic relationships between words and are capable of efficiently processing large-scale corpora. Additionally, \cite{Ye_2023_Transformer-Based} leverages the FastText method, which introduces n-gram features for learning features.

Deep learning methods can learn complex relationships and patterns in data, extracting more informative features. \cite{Rakesh_2019_Linked} employs an encoder-decoder structure to learn data features. For image data, \cite{McAuley_2015_Image-Based} uses convolutional methods for feature learning. The language model BERT (Bidirectional Encoder Representations from Transformers) can utilize contextual information in a sentence to learn semantic representations of words, better capturing the relationships between words. Compared to traditional context-independent language models such as word2vec, BERT is more suitable for natural language processing tasks that require considering contextual relevance. Additionally, BERT can be fine-tuned for specific tasks, allowing for post-training fine-tuning to optimize its performance \cite{Jian_2022_Multi-task, Li__Food, Fatemi_2023_Learning}.

Graph embedding methods can better capture the relationships between data and provide more comprehensive information. \cite{Chen_2023_Enhanced} selects PinSage \cite{Ying2018GraphCN} as the foundational model for embedding data. Additionally, \cite{Yang_2022_Inferring} utilizes relationship representation methods from knowledge graphs to process structured data, employing the TransE \cite{Bordes2013TranslatingEF} model for data embedding. Furthermore, \cite{Dai_2022_Decomposing, Ye_2023_Transformer-Based} employ LightGCN to learn from graph data.

\subsection{Challenges and Issues}
Alternative reasoning requires the ability to model and understand the real world, often involving the interaction of multiple factors. Therefore, alternative reasoning needs to effectively model and infer these complex relationships.

\subsubsection{Main Existing Problems}

The task of alternative reasoning not only involves proposing a new solution but also faces a series of challenges and problems that need to be overcome.

\textbf{Interpretability:} Interpretability is an important issue in relation to alternative reasoning. Due to the complex models and algorithms used in alternative reasoning, the decision-making process is often difficult to interpret. This limits the understanding of the reasoning results and trust in the outcomes. In \cite{Yang_2022_Inferring}, action paths from reinforcement learning are used as the basis for finding alternatives. \cite{Zhou_2022_Decoupled, Wang_2018_Path-constrained} represent the relationships and influences between different elements through graph structures, generating highly interpretable alternative solutions. \cite{Chen_2020_Try} uses key attributes as the foundation for alternative recommendations, selecting recommended item $v_j$ based on the performance of key attributes. Graph structure models have outstanding interpretability and comprehensibility in relation reasoning compared to other deep learning models. They can better understand and illustrate the connections between data, and have clear advantages in interpreting model predictions and optimizing models.

\textbf{Data Sparsity:} Sparse interaction data between users and items can lead to the issue of data sparsity. The lack of sufficient information to accurately infer user interests and item relevance poses challenges to the performance and accuracy of the model. In alternative reasoning, two common methods used to alleviate data sparsity are enhancing data representation or augmenting the dataset.

\cite{Yang_2022_Inferring} enriches the features of products using product description information as the reward function in reinforcement learning to alleviate data sparsity. \cite{Li__Food} enriches representations by pre-training the BERT model. Path representation in graph structures describes the connections between nodes through intermediate nodes, further enhancing the relationships and mutual influences between nodes. Studies such as \cite{Zhou_2022_Decoupled, Wang_2018_Path-constrained, Yang_2022_Inferring, Liu_2022_Item} expand the dataset through two-hop/ multi-hop paths of nodes. \cite{Ruiz_2019_SHOPPER} additionally introduces an indicator of interchangeability, measuring the probability of substitutability of products by calculating the similarity of product distributions.

\textbf{Personalization:} Personalization refers to how to provide customized recommendations and services based on the specific needs, interests, and behaviors of users. In the context of alternative reasoning and recommendation for products, \cite{Chen_2020_Try, Rakesh_2019_Linked} combine collaborative filtering methods to model user preferences. The study in \cite{McAuley_2015_Image-Based} extracts user preferences from user comments. Research \cite{Ruiz_2019_SHOPPER} incorporates the sequential choices of customers as a function of latent attributes and interaction coefficients into the calculation of substitutability probabilities. In the food domain, where there is no user information, but customization requirements exist for recipe substitutions, \cite{Fatemi_2023_Learning, Li__Food} provide personalized alternative recommendations based on recipe context and editing for new recipes.

\textbf{Cold Start:} The cold start problem refers to the situation where a new system or a new item introduced into the system cannot effectively provide personalized recommendations due to a lack of specific user data or item information. In addition to inferring ingredients based on recipes, \cite{Li__Food} also provides generalized ingredient inference for general substitution purposes, addressing the cold start issue. Studies such as \cite{Zhang_2022_Learning, Rakesh_2019_Linked, Wang_2018_Path-constrained, McAuley_2015_Inferring} tackle the cold start problem when new items are added to the system by using data such as titles and product descriptions for alternative reasoning, as there is a lack of user preference information for these new items. \cite{McAuley_2015_Image-Based} utilizes image data and a predictor based on visual features to perform alternative reasoning even in the absence of user preference information.

\textbf{Relationship Decoupling:} In the substitution reasoning of products, using a heterogeneous graph to model products and their relationships is a common approach that facilitates handling complementary relationships between products. Different types of relationships may mutually influence each other, leading to inaccurate reasoning results. Studies such as \cite{Zhang_2022_Learning, Zhou_2022_Decoupled, Liu_2020_Decoupled, McAuley_2015_Inferring} have constructed different subgraphs for different relationships and conducted learning on each subgraph separately. Research such as \cite{Dai_2022_Decomposing} has constructed a dependency graph for enterprises, where complementary and substitute relationships can be seen as two directions in the enterprise dependency network, and learning is conducted separately for different directions. Decoupling also provides better flexibility and scalability. \cite{McAuley_2015_Inferring} learns a set of parameters for different relationship subgraphs, making it easier to add new types of relationships or properties without causing unnecessary interference or modification to other parts of the system. Additionally, certain relationships may have interactive influences, where different relationships may depend on or interact with each other, such as the coupling of substitute and complementary relationships mentioned in 2.2.1.. By constructing subgraphs for different relationships and conducting independent learning, as well as facilitating interactive learning between the subgraphs, the complexity and interdependence of different relationship types can be addressed.

\section{SUBSTITUTE REASONING}
In the process of substitute reasoning, the primary step is to represent the data and features in a computationally understandable manner. Subsequently, the focus is on studying substitute relationships through learning, thus reasoning about substitute relationships. Finally, these independent components are integrated into a complete model and trained and optimized. The table \ref{3_tab} displays the reasoning process in substitute inference. 

\begin{table*}[]
	\centering
	\resizebox{\linewidth}{!}{
		\begin{tabular}{cccccccccc}
			\toprule
			& \multicolumn{3}{c}{}                                                         &                                     & \multicolumn{2}{c}{}                                       & \multicolumn{3}{c}{}                                                                         \\
			& \multicolumn{3}{c}{}                                                         &                                     & \multicolumn{2}{c}{}                                       & \multicolumn{3}{c}{}                                                                         \\
			& \multicolumn{3}{c}{\multirow{-3}{*}{Feature   Representation}}               &                                     & \multicolumn{2}{c}{\multirow{-3}{*}{Relation   Reasoning}} & \multicolumn{3}{c}{\multirow{-3}{*}{Model Training Optimization}}                                                 \\
			\multirow{-4}{*}{Ref} & NLP & Graph Embedding & loss                    & \multirow{-4}{*}{\makecell{ Relation \\ Learning } } &  \makecell{ Substitute \\ Reasoning }     & loss                               & global loss               & Training Optimizer & Negative Sampling                           \\\midrule
			\cite{Fatemi_2023_Learning}                  &    & GNN             & cross-entropy loss      & \checkmark                                   & \checkmark                     &  \makecell{ self-supervised \\ contrastive   loss } &                           & Adam               &                                             \\
			\cite{Ye_2023_Transformer-Based}                      &      &                 &                         & \checkmark                                   & \checkmark                     &                                    &                           & AdamW              & \makecell{ adding negative signs \\ to negative   samples } \\
			\cite{Li__Food}                       &\checkmark &                 & cross-entropy loss      &                                     & \checkmark                     &                                    &                           &                    &                                             \\
			\cite{Lawrynowicz_2022_Food}                       &                                  &                 &                         &                                     &                       &                                    &                           &                    &                                             \\
			\cite{Zhang_2022_Learning}                       &                                  & \checkmark               &                         &                                     & \checkmark                     &                                    & BPR loss                  & Adam               &                                             \\
			\cite{Zhou_2022_Decoupled}                      &                                  & GAN             &                         & \checkmark                                   &                       &                                    & cross-entropy loss        & RSGD               &                                             \\
			\cite{Jian_2022_Multi-task}                      &                                  & GNN             &                         &                                     &                       &                                    & LaAP loss                 & Adam               &                                             \\
			\cite{Dai_2022_Decomposing}                     &                                  & GCN             &                         & \checkmark                                   &                       & cross-entropy loss                 &                           & Adam               &                                             \\
			\cite{Yang_2022_Inferring}                      & \checkmark                                & KGE             &                         &                                     & \checkmark                     & cross-entropy loss                 &                           &                    &                                             \\
			\cite{Pellegrini_2021_Exploiting}                      & \checkmark                                &                 &                         &                                     &                       &                                    &                           & AdamW              &                                             \\
			\cite{Shirai_2021_Identifying}                     & \checkmark                                &                 &                         &                                     &                       &                                    &                           &                    &                                             \\
			\cite{Chen_2020_Try}                      &                                  & \checkmark               &                         & \checkmark                                   & \checkmark                     & BPR loss                           &                           & Adam               &                                             \\
			\cite{Liu_2020_Decoupled}                     &                                  & \checkmark               &                         & \checkmark                                   & \checkmark                     &                                    & graph-based loss          & Adam               &                                             \\
			\cite{Chen_2020_Studying}                      & \checkmark                                &                 &                         &                                     & \checkmark                     &                                    &                           &                    & probability                                 \\
			\cite{Zhang_2019_Identifying}                     & \checkmark                                &                 &                         &                                     & \checkmark                     &  \makecell{ categorical \\ cross-entropy loss   }     &                           & Adam               &                                             \\
			\cite{Shen_2019_KGDDS}                      &                                  &                 &                         & \checkmark                                   &                       &                                    &                           &                    &                                             \\
			\cite{Zhang_2019_Inferring}                      &                                  &                 &                         & \checkmark                                   & \checkmark                     &                                    &                           &                    &                                             \\
			\cite{Rakesh_2019_Linked}                     &                                  &                 &                         & \checkmark                                   & \checkmark                     &                                    &                           &                    &                                             \\
			\cite{Wang_2018_Path-constrained}                      & \checkmark                                &                 & logistic loss           & \checkmark                                   & \checkmark                     &                                    &                           & SGD                &                                             \\
			\cite{Ruiz_2019_SHOPPER}                      &                                  &                 &                         & \checkmark                                   & \checkmark                     &                                    &                           &                    &                                             \\
			\cite{McAuley_2015_Inferring}                      & \checkmark                                &                 &                         & \checkmark                                   &                       &                                    &                           & L-BFGS             &                                             \\
			\cite{McAuley_2015_Image-Based}                     &                                  &                 &                         &                                     & \checkmark                     &                                    &                           & L-BFGS             &                                             \\
			\cite{Boscarino_2014_Automatic}                      & \checkmark                                &                 &                         & \checkmark                                   &                       &                                    &                           &                    &                                             \\
			\cite{Chen_2023_Enhanced}                     &                                  & \checkmark               & max-margin ranking loss & \checkmark                                   & \checkmark                     &                                    &                           & Adam               &                                             \\
			\cite{Liu_2022_Item}                      &                                  & \checkmark               &                         & \checkmark                                   & \checkmark                     &                                    & \makecell{ sum of \\ cross-entropy loss   } & Adam               &                \\\bottomrule                                
		\end{tabular}
	}
\caption{Inference Process in Substitute Reasoning}
	\label{3_tab}
\end{table*}

\subsection{Feature Representation Method}
The feature representation method refers to extracting features from the raw data in order to better analyze, and predict the data. 

As mentioned in the literature \cite{Boscarino_2014_Automatic,McAuley_2015_Inferring,Zhang_2019_Identifying}, the LDA method was used for dimensionality reduction and embedding of topic distribution. In the study \cite{Zhang_2019_Identifying}, a different approach was employed for similarity computation, using the outer product of the two vectors to explore all associations between two items (i.e., $\theta_i$ and $\theta_j$), rather than the inner product commonly used in most similar calculations.
In the study, a feature set comprising seven important non-textual factors was developed, such as the number of comments, average ratings, variance, and so on. The textual and non-textual features were eventually concatenated into an (4$K^2$+7)×1 vector representation of the product.The study \cite{McAuley_2015_Inferring} suggests that similar products typically have similar review texts. It uses logistic regression (sigmoid function) to predict the connections between products. Furthermore, the study also takes into account the directional nature of relationships and describes asymmetry in direction by modeling the differences between product topic vectors.
\begin{footnotesize}
\begin{equation}
	\begin{aligned} & L(y, \mathcal{T} \mid \beta, \eta, \theta, \phi, z)= \\ & \overbrace{\prod_{(i, j) \in \mathcal{E}} F_\beta^{\leftrightarrow}\left(\psi_\theta(i, j)\right) F_\eta \rightarrow\left(\varphi_\theta(i, j)\right)\left(1-F_\eta\left(\varphi_\theta(j, i)\right)\right)}^{\text {positive relations }\left(F^{\leftrightarrow}\right) \text { and their direction of flow }\left(F^{\rightarrow}\right)} \\ & \underbrace{\prod_{(i, j) \in \overline{\mathcal{E}}}\left(1-F_\beta^{\leftrightarrow}\left(\psi_\theta(i, j)\right)\right)}_{\text {non-relations }} \underbrace{\prod_{d \in \mathcal{T}} \prod_{j=1}^{N_d} \theta_{z_{d, j}} \phi_{z_{d, j}, w_{d, j}}}_{\text {corpus likelihood }} &
	\end{aligned}
\end{equation}
\end{footnotesize}
In \cite{Wang_2018_Path-constrained}, the method for link prediction also utilizes the sigmoid function and involves two types of vectors: the target vector V and the context vector $V^{\prime}$, which are used to capture different semantics of products in directed relationships.

Studies such as \cite{McAuley_2015_Image-Based,Fatemi_2023_Learning,Pellegrini_2021_Exploiting,Jian_2022_Multi-task} have utilized image data. In the study by \cite{McAuley_2015_Image-Based}, they used the Mahalanobis distance based on the projection of image feature vectors to measure the similarity between products. In the study \cite{Pellegrini_2021_Exploiting}, ImageNet \cite{Deng2009ImageNetAL} was used to embed a portion of the images in the data. For the textual part, two embedding methods were proposed: (1) utilizing a variation of CBOW from Word2vec for computation, and (2) pretraining with an extended vocabulary from ingredient data using BERT and performing vector dimensionality reduction with PCA. In \cite{Fatemi_2023_Learning}, ViT \cite{Dosovitskiy2020AnII} was used as the encoder for images. As for textual data, graph neural networks are the most commonly used method for processing graph-structured data. The study utilized multiple Graph Isomorphism Network (GIN) \cite{Xu2018HowPA} layers as the component encoders, with each GIN layer defined as follows:

\begin{footnotesize}
\begin{equation}
	\mathbf{h}_v^{(l)}=f^{(l)}\left(\left(1+\epsilon^{(l)}\right) \mathbf{h}_v^{(l-1)}+\sum_{u \in \mathcal{N}(v)}\left(e_{v u} \mathbf{h}_u^{(l-1)}\right) ; \theta^{(l)}\right)
\end{equation}
\end{footnotesize}

In the study \cite{Dai_2022_Decomposing}, LightGCN \cite{He2020LightGCNSA} was used to embed graph data. LightGCN is a lightweight graph neural network model that was used to embed the dependency relationship network of enterprises.
In the study \cite{Zhou_2022_Decoupled}, the data was embedded into hyperbolic space, and an attention mechanism was used for information aggregation. Unlike traditional Euclidean space, hyperbolic space can better handle data with complex structures and non-linear relationships.

In the study \cite{Zhang_2022_Learning}, the aim was to obtain disentangled representations of specific relationships for each product. Two graphs, Gs and Gc, were constructed based on known substitutive and complementary relationships. Complex dependency relationships were updated by combining data from different relationships in the representation updating phase.
\begin{equation}
	\begin{aligned}
		\bar{h}_{s, p}^k = & \sum_{j \in \mathcal{N}_s(p)} \hat{A}_s[p, j] \cdot h_{s, j}^{k-1} \\
		& +\tanh \left(\hat{A}_s[p, j] \cdot h_{s, j}^{k-1}\right) \odot\left(h_{s, j}^{k-1}+r_s\right)\\
		& +\left(\hat{A}_s[p, j] \cdot h_{s, j}^{k-1}+r_s\right) \odot \tanh \left(h_{s, j}^{k-1}\right)
	\end{aligned}
\end{equation}
\begin{equation}
	\boldsymbol{h}_{s, p}^k=\left(1-\alpha_{s, p}^k\right) \cdot \overline{\boldsymbol{h}}_{s, p}^k+\alpha_{s, p}^k \cdot T_{c->s}\left(\overline{\boldsymbol{h}}_{c, p}^k ; \theta_s^T\right)
\end{equation}

In the study \cite{Li__Food}, knowledge graphs, as a special type of graph structure, were embedded using the pre-trained language model BERT. The research retrained BERT on two tasks: predicting substituent composition and assessing the plausibility of triplets within the knowledge graph.

In addition to graph neural networks, the study \cite{Yang_2022_Inferring} employed reinforcement learning methods using TransE \cite{Bordes2013TranslatingEF} for embedding representations. The research formulated the inference of product relationships using structured data as a Markov decision process, with the following process:

In the study, non-structured data was embedded using TF-IDF and Doc2vec \cite{Le2014DistributedRO}. In the research \cite{Chen_2020_Studying}, a Product2Vec model was created, which utilized basket composition information to transform products into low-dimensional vectors with continuous elements.
\begin{equation}
	\mathcal{V}, \mathcal{V}^{\prime}=\underset{\mathcal{V}, \mathcal{V}^{\prime}}{\arg \max } \sum_{b \in B} \sum_{s_i \in b} \sum_{-c \leq j \leq c} \log \mathbb{P}\left(s_{i+j} \mid s_i ; \mathcal{V}, \mathcal{V}^{\prime}\right)
\end{equation}

In \cite{Ruiz_2019_SHOPPER}, a logarithmic bilinear form is employed, where the probability of a customer selecting item c at step i depends on the latent features of item c and the user preference $\theta_u$.
\begin{equation}
	\begin{aligned}
	\operatorname{Prob}(\text { item } c \mid \text { items in basket }) \propto \\
	\exp \left\{\theta_u^{\top} \alpha_c+\rho_c^{\top}\left(\frac{1}{i-1} \sum_{j=1}^{i-1} \alpha_{y_j}\right)\right\}
	\end{aligned}
\end{equation}

In the study \cite{Chen_2020_Try}, collaborative filtering is applied separately to users X and items Y, and then combined. Traditional matrix factorization-based collaborative filtering methods have limited expressive power. In the research, residual feedforward networks are used to model pairwise attributes.

LDA and TF-IDF are commonly used methods in text analysis and topic modeling, but they disregard the specific meanings and semantic information of words. Word2Vec is a deep learning model used for learning word vector representations, which captures the semantic relationships between words but only captures local context dependencies. In contrast, BERT \cite{Devlin2019BERTPO} can capture longer-range contextual information but requires a large amount of text data for pre-training. The choice of appropriate representation method depends on the characteristics of the data, such as a large amount of lengthy text data being more suitable for BERT, while less and shorter text can be considered for Word2Vec.

For structured data such as graph data, relationship data, and knowledge graphs, more attention is focused on the structure of the data rather than textual semantics. By employing comprehensive representation learning methods, it becomes possible to consider information related to structure, relationships, and semantics simultaneously, thereby improving the performance and effectiveness of structured data analysis tasks.

\subsection{Substitute Learning and Inference}
Substitute learning and inference involve using the learned features to infer other possible facts or relationships.

\subsubsection{Relation Learning}
Relation learning refers to the process of learning and understanding the relationships between entities from given data. These relationships can include interactions, dependencies, associations, and more. Most often, relation learning specifically refers to the learning of alternative relationships, while some models also encompass complementary relationships. This text will only focus on the learning and inference of alternative relationships. 

In \cite{Boscarino_2014_Automatic}, a set of alternative rules was manually derived as the basis for alternative reasoning. In \cite{Ruiz_2019_SHOPPER}, it is argued that the probability of selecting item c depends on its features, so additional influencing factors (preferences, price factors, seasonal factors, popularity) were included for evaluation. 
In \cite{Rakesh_2019_Linked}, the authors linked two Variational Autoencoders (VAEs) \cite{Kingma2013AutoEncodingVB}, with one serving as the encoder to sample features and the other serving as the decoder to learn relational features. Finally, the classifier part is implemented through the softmax function. The implemented function is as follows, where $D_a$ and $D_b$ represent the KL divergence terms for items a and b, V represents the total number of data points (or items), and M represents a random sample set extracted from V.
\begin{equation}
	\begin{aligned}
		\frac{V}{M} \sum_{i=1}^M & \left[-\frac{1}{2 \sigma^2}\left(x_i^a-\mu_\phi\left(x_i^a\right)\right)^2-\mathcal{D}^a\right. \\
		& -\frac{1}{2 \sigma^2}\left(x_i^b-\mu_\phi\left(x_i^b\right)\right)^2-\mathcal{D}^b \\
		& \left. +\log \left(\frac{e^{\theta^{(y) T} \mathbf{z}}}{\sum_{j=1}^Y e^{\theta^{(j) T} \mathbf{z}}}\right)\right]
	\end{aligned}
\end{equation}

In the study \cite{Zhang_2019_Inferring}, it is argued that substitutable goods are not commonly purchased together, implying that there is no co-purchasing relationship between substitute products. Based on this, the authors propose a method to enhance second-order proximity and reduce first-order proximity. They use Autoencoders (AE) to embed the entire co-purchasing graph of goods, learn information from the neighborhood of nodes, and add constraints to the connections between nodes.
\begin{equation}
	\begin{aligned}
		& \mathbf{y}_i^{(1)}=\sigma\left(\mathbf{W}^{(1)} \mathbf{x}_i+\mathbf{b}^{(1)}\right) \\
		& \mathbf{y}_i^{(k)}=\sigma\left(\mathbf{W}^{(k)} \mathbf{y}_i^{(k-1)}+\mathbf{b}^{(k)}\right), k=2, \ldots, K
	\end{aligned}
\end{equation}
\begin{equation}
	\begin{aligned}
		&d\left(\mathbf{y}_i^{(K)}, \mathbf{y}_j^{(K)}\right) < d\left(\mathbf{y}_i^{(K)}, \mathbf{y}_t^{(K)}\right), \\
		&\forall v_i \in V, \forall v_j \in V_i^{\text{sub}}, \forall v_t \in V_i^{\text{com}}
	\end{aligned}
\end{equation}

For cases with multiple relationships, studies such as \cite{McAuley_2015_Inferring,Wang_2018_Path-constrained,Liu_2020_Decoupled,Dai_2022_Decomposing}, and \cite{Zhou_2022_Decoupled} focus on learning for different relationships independently. In \cite{McAuley_2015_Inferring}, a predictor is trained for each subgraph formed by individual relationships. In the research by \cite{Chen_2023_Enhanced}, the initial single-relation features are integrated, and then the method from \cite{Ma2018ModelingTR} is used for structural integration, exploring the mutual influence between different types of relation neighbors, leading to improved model accuracy through knowledge transfer between different relationships. Additionally, in \cite{Wang_2018_Path-constrained}, vectors are projected into different relational spaces.
\begin{equation}
	v_{r, i}=v_i+\beta_r \odot v_i, v_{r, i}^{\prime}=v_i^{\prime}+\beta_r \odot v_i^{\prime}
\end{equation}
The study \cite{Liu_2020_Decoupled} suggests that there may be inherent mutual influences between relationships and proposes a Co-attention neighborhood aggregation strategy, utilizing \cite{Xiong2016DynamicCN}, to learn the semantics of different relationships by replacing parts of the subgraph's structure.
\begin{equation}
	\begin{aligned}
		& \widetilde{\mathbf{H}}_c=\left[\mathbf{H}_c ; \mathbf{A}_c \mathbf{H}_s\right] \\
		& \widetilde{\mathbf{H}}_s=\left[\mathbf{H}_s ; \mathbf{A}_s \widetilde{\mathbf{H}}_c\right]
	\end{aligned}
\end{equation}
The study by \cite{Dai_2022_Decomposing} suggests that firms in interfirm networks rely on substitute firms to share similar contexts in the parallel direction, and they use direction-based graph convolutional methods to capture this substitutive relationship.
\begin{equation}
	\begin{aligned}
		& P_{i, \uparrow}^{(l+1)}=\sigma\left(\sum_r \sum_{j \in \mathcal{N}_{i, \uparrow}^r} \frac{1}{c_{i, r, \uparrow}^{(l)}} W_{r, \uparrow}^{(l)} P_j^{(l)}\right), \\
		& P_{i, \downarrow}^{(l+1)}=\sigma\left(\sum_r \sum_{j \in \mathcal{N}_{i, \downarrow}^r} \frac{1}{c_{i, r, \downarrow}^{(l)}} W_{r, \downarrow}^{(l)} P_j^{(l)}\right)
	\end{aligned}
\end{equation}
In \cite{Zhou_2022_Decoupled} proposed DHGAN model, which decouples and embeds relationships in hyperbolic space. To propagate the mutual influence between two different relationships, two rounds of aggregation operations are performed, as illustrated in the model diagram in Fig. \ref{4-1-4_img}.
\begin{equation}
	\text { target node } \stackrel{s}{\longleftarrow} \text { c neighbors } \stackrel{s}{\dashleftarrow} \text { s neighbors, }
\end{equation}
\begin{figure*}[htbp]
	\centering
	\includegraphics[width=0.75\textwidth]{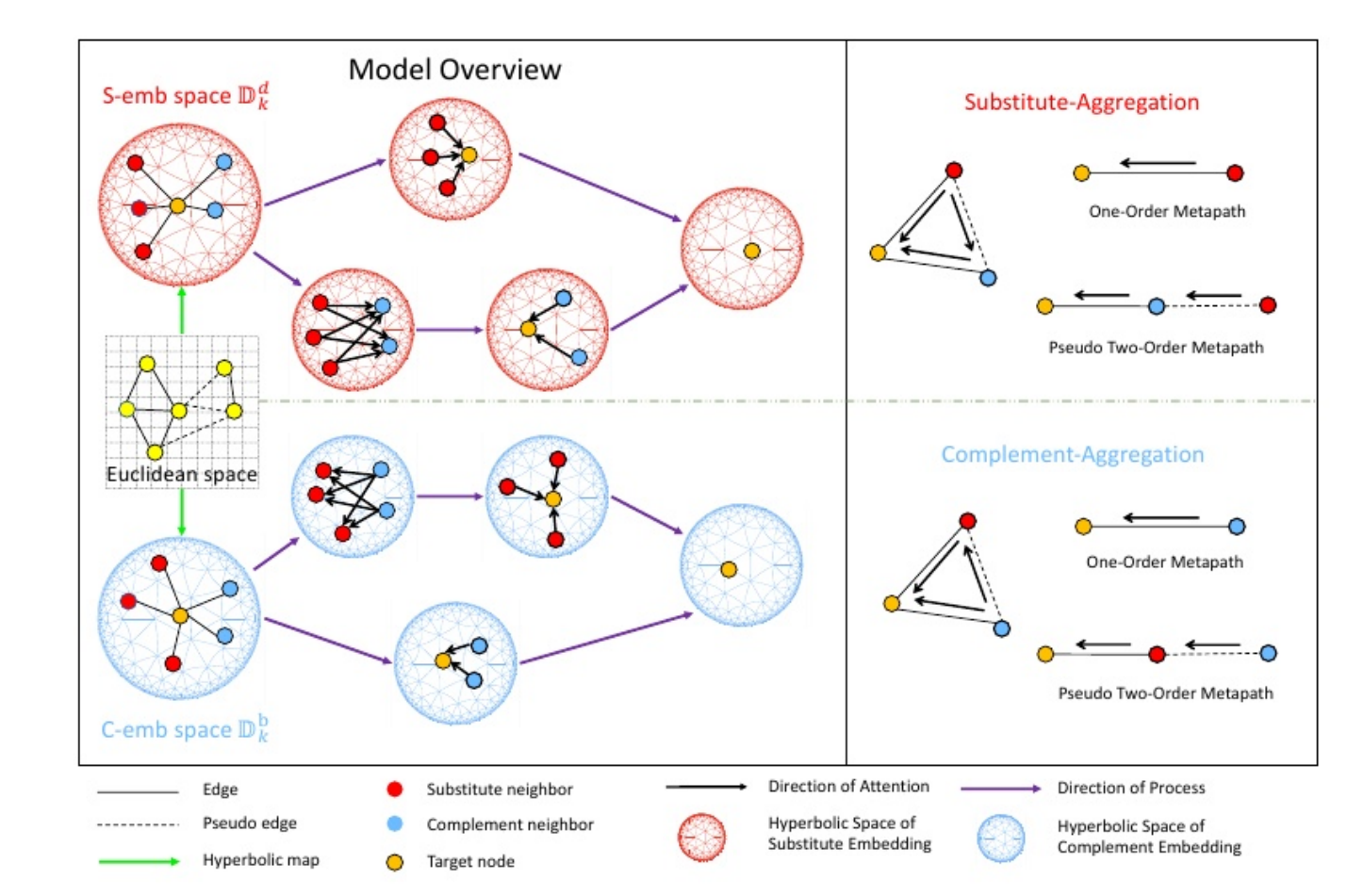}	
	\caption{DHGAN\protect}    
	\label{4-1-4_img}
\end{figure*}

In \cite{Fatemi_2023_Learning}, the authors utilize the pre-trained CLIP model \cite{Radford2021LearningTV} to learn the context of recipes. On the other hand, \cite{Ye_2023_Transformer-Based} suggests that the elasticity effect of purchase rate (PR) reflects the substitutability of products. They treat the purchase rate as a label and apply a logarithmic transformation to alleviate the issue of a long-tail 

\subsubsection{Substitutive Reasoning}
Once the model has learned the characteristics of substitutive relationships, it can include predicting missing data and filling in unknown relationships through reasoning or modeling.

For example, studies such as \cite{Boscarino_2014_Automatic,McAuley_2015_Image-Based,Zhang_2019_Inferring,Chen_2020_Studying} have measured the strength of relationships between items by calculating their similarity. According to the study \cite{Chen_2020_Studying}, if both A and B have similar interactions with other products, they are considered interchangeable.
\begin{footnotesize}
\begin{equation}
	\begin{aligned}
		E_{A B} = & -\frac{1}{2}[KL(p(\cdot \mid A) \| p(\cdot \mid B)) + KL(p(\cdot \mid B) \| p(\cdot \mid A))] \\
		= & -\frac{1}{2} \sum_{k \neq A, B} \left[p(k \mid A) \cdot \log \left(\frac{p(k \mid A)}{p(k \mid B)}\right) \right.\\
		& \left. + p(k \mid B) \cdot \log \left(\frac{p(k \mid B)}{p(k \mid A)}\right)\right]
	\end{aligned}
\end{equation}
\end{footnotesize}

In the study \cite{Zhang_2019_Inferring}, in order to further capture the semantic similarity between substitute items, the comparison of their semantics is based on the precise positions of two distinct products in the product category tree (see Figure \ref{product-tree_img}).
\begin{figure}[htbp]
	\centering
	\includegraphics[width=0.5\textwidth]{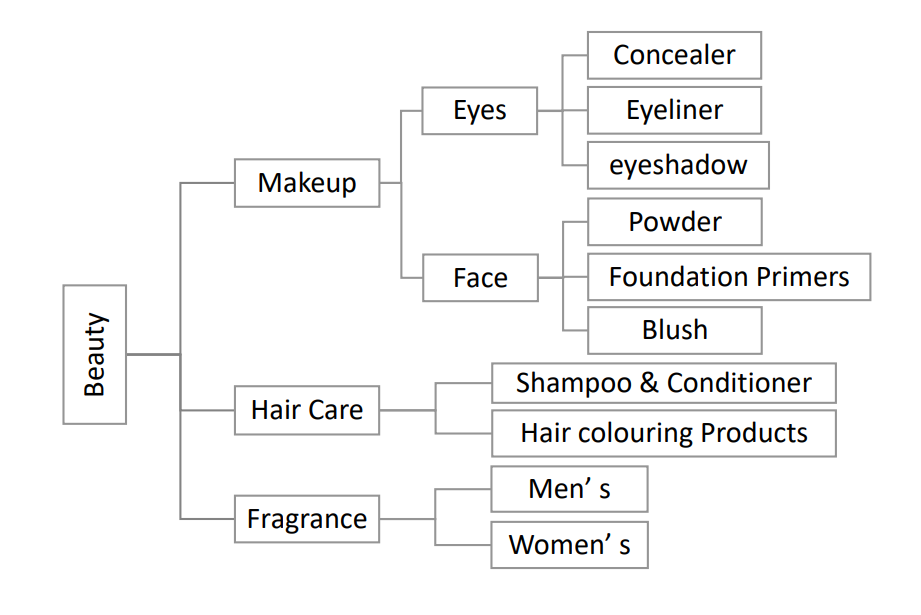}	
	\caption{Product category tree\protect}    
	\label{product-tree_img}
\end{figure}

According to \cite{McAuley_2015_Inferring}, substitution between products only occurs when they belong to similar categories. They use the product category tree and the path where the products are located as the basis for convergence in substitution inference. Similar studies include \cite{Lawrynowicz_2022_Food,Yang_2022_Inferring,Wang_2018_Path-constrained}, which also seek substitute items within similar categories.

In the substitution inference process, \cite{Wang_2018_Path-constrained} introduces two constraints: one is the product category constraint, and the other is the extension of the relationship between products through two-hop paths. This method helps alleviate the sparsity issue in product relationships.

The study\cite{Ruiz_2019_SHOPPER,Rakesh_2019_Linked,Chen_2020_Try,Zhang_2022_Learning,Dai_2022_Decomposing} combine substitution inference with personalized recommendations. In the research by \cite{Rakesh_2019_Linked,Chen_2020_Try,Dai_2022_Decomposing}, user information is integrated in the form of collaborative filtering. Research \cite{Ruiz_2019_SHOPPER} introduces the element of ``forward thinking", where each customer's choice can change the purchase probability of substitute products. Additionally, \cite{Chen_2020_Try} proposes a variant of BPR \cite{Rendle2009BPRBP} to unify substitution and personalization.\cite{Zhang_2022_Learning} leverages different temporal patterns of sequential behaviors to understand user preferences in browsing and purchasing products. They propose a kernel transformer network for analysis in this context.

In the study by \cite{Zhang_2019_Identifying}, a feed-forward neural network is used to determine substitution relationships. However, it may suffer from low interpretability. On the other hand, \cite{Ye_2023_Transformer-Based} combines Gradient Boosting Decision Trees (GBDT) with a transformer-based model called XLM-R \cite{Conneau2019UnsupervisedCR} for text sequence processing. In the research conducted by \cite{Li__Food}, text adversarial attacks are employed. The study proposes three strategies for these attacks, namely recipe editing, personalized ingredient substitution, and universal ingredient replacement.

In the study by \cite{Liu_2022_Item}, after obtaining embeddings, they propose an outer product layer to reconstruct a given graph structure. The outer product operation, represented as $h_v^{(L)} \otimes h_w^{(L)}$, is non-commutative, meaning that $\left[h_v^{(L)}, h_w^{(L)}\right]$ and $\left[h_w^{(L)}, h_v^{(L)}\right]$ yield different results. Therefore, the direction can be utilized for prediction purposes.

\textbf{Loss Calculation:} 
In the study by \cite{Fatemi_2023_Learning}, a self-supervised contrastive loss function is employed \cite{Chen2020ASF, Kadlec2017KnowledgeBC}. On the other hand, \cite{Zhang_2019_Identifying, Dai_2022_Decomposing, Yang_2022_Inferring} utilize cross-entropy loss functions, with \cite{Zhang_2019_Identifying} employing categorical cross-entropy.

In \cite{Zhang_2019_Inferring}, a penalty is applied when two substitutable products under similar categories in the category tree are mapped far apart in the embedding space based on their positions in the category tree.The model in \cite{Zhou_2022_Decoupled} applies graph-based loss for each sub-GCN, inferring different relationships, and the final loss function can be represented as a multi-task loss.

In \cite{Yang_2022_Inferring}, the loss is determined based on a reward function:
\begin{equation}
	\begin{aligned}
		\mathcal{L} = & \sum_{i, j \in \varepsilon_P} y_{i, j} \cdot \log \operatorname{MFI}(i, j) \\
		& + \sum_{i, j \in \bar{\varepsilon}_P} \left(1 - y_{i, j}\right) \cdot \log (1 - \operatorname{MFI}(i, j))
	\end{aligned}
\end{equation}

\subsubsection{Simplified Inference Methods}

In this survey, we divide substitution reasoning into four parts: feature representation, relationship learning, substitution reasoning, and integrated training. However, some studies focus only on one or two of these parts, making them more streamlined and lightweight compared to the aforementioned substitution reasoning models. Therefore, in the following content, we will discuss these methods.

In \cite{Shen_2019_KGDDS}, a large-scale antibiotic-related knowledge graph is constructed, utilizing TF-IDF embeddings and feed-forward layers for dimensionality reduction. All entity descriptions in the knowledge graph are extracted from Drugbank, and an attention mechanism is used to allocate feature weights. This approach aims to understand drug properties from the perspective of drug similarity for substitution.

In \cite{Pellegrini_2021_Exploiting}, text embedding methods such as Word2Vec and BERT are utilized to train Food2Vec and FoodBERT models on a recipe dataset (as shown in Figure \ref{5-3-2_img}). The study also introduces the concept of multimodality by integrated use of textual and visual information.

In \cite{Shirai_2021_Identifying}, multiple scoring criteria are combined to develop a heuristic method called DIISH for identifying ingredient substitutability. The method incorporates the following four components: Obtaining latent semantic information using word embedding models; Calculating Positive Pointwise Mutual Information (PPMI) by considering contextual information; Computing co-occurrence substitutability scores between ingredients A and B across different recipes; Determining the similarity between the two ingredients and the recipes. By incorporating scores from these four components, the DIISH method provides insights into ingredient substitutability.

In \cite{Jian_2022_Multi-task}, three different customer behavior data sources, namely CSS, search logs, and OOS, were selected. Image and title text features were extracted from each product as input features. The M-HetSage model (as shown in Figure 5-3-2) was proposed, which combines the loss functions from different datasets and weights them to unify several different tasks within one architecture.
\begin{figure}[]
	\centering
	\includegraphics[width=0.48\textwidth]{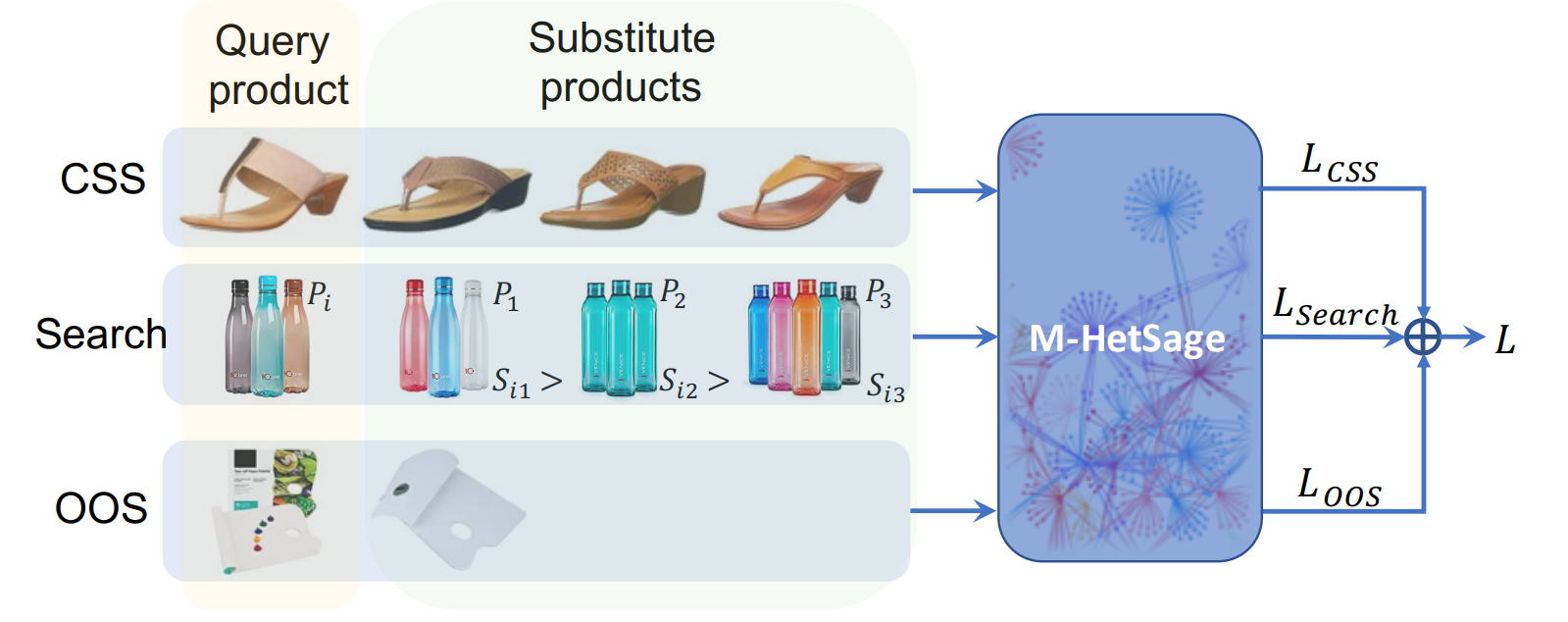}	
	\caption{M-HetSage\protect}    
	\label{5-3-2_img}
\end{figure}
Furthermore, a LaAP loss was proposed, which directly optimizes the target ranking metric mAP and is equipped with a list-wise attention mechanism.
\begin{equation}
	\begin{aligned}
		& A P\left(\mathbf{Z}_i, \mathrm{Y}_i .\right)=\frac{1}{N_i^{+}} \sum_{k=1}^{N_i} C_k\left(\mathbf{Z}_i, \mathrm{Y}_i .\right) r_k\left(\mathrm{Z}_i, \mathrm{Y}_i \cdot\right) \\
		& q A P\left(\mathbf{Z}_i, \mathbf{Y}_i .\right)=\frac{1}{N_i^{+}} \sum_{m=1}^M \hat{C}_m\left(\mathbf{Z}_i, \mathbf{Y}_i .\right) \hat{r}_m\left(\mathbf{Z}_i, \mathbf{Y}_i .\right) \\
		& \operatorname{LaAP}\left(\mathbf{Z}_i, \mathrm{Y}_i .\right)=\frac{1}{B} \sum_{i=1}^B w_i \cdot q A P\left(\mathbf{Z}_i, \mathbf{Y}_i \cdot\right), \\
		&
	\end{aligned}
\end{equation}

\subsection{Model Training}
The independently described components are organically integrated into a complete model, with each part playing a crucial role in the integration process, providing different optimization contributions to the performance of the model.

\subsubsection{Integration of Components:} The method proposed by \cite{Fatemi_2023_Learning} decouples ingredient substitution reasoning and recipe editing, by first obtaining the optimal ingredient replacement and then combining it with image information to decode the recipe for recipe editing. Different weights are assigned to each module, as described in \cite{Wang_2018_Path-constrained,Zhang_2019_Inferring,Zhou_2022_Decoupled,Dai_2022_Decomposing}, and the scores are summed to obtain the final result. Attention mechanisms are employed in \cite{Zhou_2022_Decoupled,Dai_2022_Decomposing} to allocate the weights. The research conducted by \cite{Chen_2020_Studying,Ruiz_2019_SHOPPER} takes into consideration factors such as price, preferences, marketing sensitivity, as well as complementary and substitution relationships, to integrate the final results.

\subsubsection{Training Optimization}
Training optimization involves multiple aspects: whether the dataset used has annotated information, the loss function used to calculate the model's loss.

\textbf{Supervised Learning:}
Currently, most machine learning applications still adopt supervised learning methods because annotated data can often provide more accurate supervision signals, resulting in better-trained models.

In the research \cite{Ye_2023_Transformer-Based} uses weakly supervised learning to improve model performance in situations where supervision signals are limited or inaccurate by supplementing them with enhanced supervision signals.

Studies such as \cite{Chen_2020_Studying,Rakesh_2019_Linked,Zhang_2019_Inferring} employ unsupervised learning to learn models by automatically discovering patterns or features from the data.

The research conducted by \cite{Fatemi_2023_Learning} utilizes self-supervised contrastive losses on textual data. It learns representations by generating fake samples and contrasting them with real samples.

\textbf{Negative Sample Sampling:}
A common method is to use random sampling, where a subset of samples is randomly selected from the dataset as negative samples. In the research conducted by \cite{Ye_2023_Transformer-Based}, increasing the negative samples can lead to negative samples showing an opposite trend or distribution to positive samples in the feature space, helping the model better distinguish between positive and negative samples. In the study by \cite{Chen_2020_Studying}, products are randomly sampled from the entire training set based on their purchase frequency distribution, minimizing the likelihood of the current product appearing with randomly chosen irrelevant products. The selection and handling methods for negative samples depend on the specific task and dataset, and need to be adjusted and experimented with based on the actual situation to obtain better model performance.

\section{EVALUATION AND DATASETS}

In this chapter, different datasets and evaluation metrics commonly used for various inference tasks are introduced, and alternative practical applications are summarized. This section can help researchers find suitable datasets and evaluation metrics to test their methods.

\subsection{Available Datasets}
\textbf{Commodity Dataset:}

\begin{itemize}
	\item[$\bullet$]Amazon Dataset: It includes reviews (ratings, text, helpfulness votes), product metadata (descriptions, category information, prices, brands, and image features), and links (also bought/also viewed charts). The complete dataset is divided into subdatasets by category, such as Amazon-Books, Amazon-Instant Videos, and Amazon-Electronics. Subdatasets within Amazon are commonly used to test the performance of user-item relationship inference and alternative recommendation methods.
	\item[$\bullet$]E-Commerce Data: It contains all the transactions of a certain company's non-physical online retail registered in the UK from December 1, 2010 to December 9, 2011. The company mainly sells unique all-weather gifts, and many of its customers are wholesalers. 
\end{itemize}

\textbf{Food Dataset:}

\begin{itemize}
	\item[$\bullet$]FoodOn: (http://foodon.org) is an alliance-driven project aimed at establishing a comprehensive and user-friendly global food ontology, accurately and consistently describing universally known foods from various cultures around the world, from farm to table. It addresses gaps in food terminology and supports the traceability of food.
	\item[$\bullet$]Recipes 1M+: A large-scale, structured corpus consisting of over one million cooking recipes and 13 million food images. It is a publicly available recipe dataset that offers alignable multimodal data.
	\item[$\bullet$]Open Food Facts: is a free, open, collaborative global food database that contains information about ingredients, allergens, nutritional values, and all the information we can find on product labels. It includes data from over 600,000 products from more than 150 countries/regions.
	\item[$\bullet$]FoodKG: FoodKG is a large-scale, unified food knowledge graph that integrates food ontology, recipes, ingredients, and nutritional data. It incorporates FoodOn into its WhatToMake ontology and includes recipe and nutrient instances extracted from Recipe1M+ as well as nutrient records from the United States Department of Agriculture (USDA). A comprehensive food knowledge graph with extensive recipe and nutritional information can support various applications such as recipe recommendation, ingredient substitution, and quality control.
\end{itemize}

\begin{table*}[]
	
	\centering
	\resizebox{\linewidth}{!}{
		\begin{tabular}{lllll}
			\toprule 
			Datasets             & Field & Links  & Size   & Refs                                                 \\\midrule
			Amazon Dataset        & E-Commerce & http://jmcauley.ucsd.edu/data/amazon/ & \begin{tabular}[c]{@{}l@{}}Ratings: 82.83 million\\      Users: 20.98 million\\      Items: 9.35 million\\      Timespan: May 1996 - July 2014\\  \end{tabular} &  \begin{tabular}[c]{@{}l@{}}\cite{Yang_2022_Inferring,Zhang_2022_Learning,Zhou_2022_Decoupled,Chen_2020_Try,Zhang_2019_Identifying,Zhang_2019_Inferring,McAuley_2015_Inferring}\\ \cite{McAuley_2015_Image-Based,Wang_2018_Path-constrained,Liu_2020_Decoupled,Rakesh_2019_Linked,Chen_2023_Enhanced,Liu_2022_Item} \end{tabular}              \\\midrule
			Online Retail & E-Commerce & https://archive.ics.uci.edu/dataset/352/online+retail & \begin{tabular}[c]{@{}l@{}}Instances:541909\\      Features:6\\      Timespan:01/12/2010 and 09/12/2011\\  \end{tabular}&\cite{Ruiz_2019_SHOPPER}\\\midrule
			FoodOn          & Food & http://foodon.org    & over 9,600 generic food product  &   \cite{Lawrynowicz_2022_Food,Lawrynowicz_2022_Food}                               \\\midrule
			Recipes 1M+     & Food & http://im2recipe.csail.mit.edu/ & \begin{tabular}[c]{@{}l@{}}structured cooking recipes :one   million \\      associated images: 13M\\  \end{tabular}  &  \cite{Pellegrini_2021_Exploiting,Fatemi_2023_Learning}                     \\\midrule
			Open Food Facts & Food & https://world.openfoodfacts.org/  & product: 3068832 &                     \\
			FoodKG          & Food & https://foodkg.github.io     & approx 63 million triples &  \cite{Li__Food}            \\\bottomrule           
		\end{tabular}
	}
\caption{Dataset}
\end{table*}

\subsection{Evaluation Criteria}
Selecting appropriate metrics to assess the performance of comparative methods is of paramount importance. Below is a summary of the evaluation criteria utilized for various tasks.

\begin{equation}
	\text { HitRate }=\frac{T P}{T P+F N}
\end{equation}

\begin{table}[h]

	\centering
	\scalebox{0.65}{ 	
		\begin{tabular}{cccccccccc}
			\toprule 
			Literature & Recall & MRR & HR & AuPRC & NDGC & mAP & Hit & Accuracy & Precision \\\midrule
			\cite{Fatemi_2023_Learning}  &        & \checkmark   &    &       &      &     & \checkmark   &          &           \\
			\cite{Ye_2023_Transformer-Based}  &        &     &    & \checkmark     & \checkmark    &     &     &          &           \\
			\cite{Li__Food}  &        & \checkmark   &    &       &      &     & \checkmark   &          &           \\
			\cite{Zhang_2022_Learning} &        &     & \checkmark  &       & \checkmark    &     &     &          &           \\
			\cite{Zhou_2022_Decoupled} &        & \checkmark   & \checkmark  &       & \checkmark    &     &     &          &           \\
			\cite{Jian_2022_Multi-task} &        &     &    &       & \checkmark    & \checkmark   &     &          &           \\
			\cite{Dai_2022_Decomposing} & \checkmark      &     &    &       &      & \checkmark   &     &          &           \\
			\cite{Yang_2022_Inferring} & \checkmark      &     & \checkmark  &       & \checkmark    &     &     &          & \checkmark         \\
			\cite{Pellegrini_2021_Exploiting} & \checkmark      &     &    &       &      &     &     &          & \checkmark         \\
			\cite{Shirai_2021_Identifying} & \checkmark      & \checkmark   &    &       &      & \checkmark   &     &          &           \\
			\cite{Chen_2020_Try} &        &     & \checkmark  &       & \checkmark    &     &     &          &           \\
			\cite{Liu_2020_Decoupled} &        & \checkmark   & \checkmark  &       & \checkmark    &     &     &          &           \\
			\cite{Chen_2020_Studying} &        &     &    &       &      &     &     &          &           \\
			\cite{Zhang_2019_Identifying} &        &     &    &       &      &     &     &          & \checkmark         \\
			\cite{Shen_2019_KGDDS} &        &     &    &       &      &     &     &          &           \\
			\cite{Zhang_2019_Inferring} &        &     &    &       &      &     & \checkmark   &          &           \\
			\cite{Rakesh_2019_Linked} &        &     &    &       &      &     &     &          & \checkmark         \\
			\cite{Wang_2018_Path-constrained} &        &     &    &       &      &     &     &          & \checkmark         \\
			\cite{Ruiz_2019_SHOPPER} &        &     &    &       &      &     &     &          &           \\
			\cite{McAuley_2015_Inferring} &        &     &    &       &      &     &     &          & \checkmark         \\
			\cite{McAuley_2015_Image-Based} &        &     &    &       &      &     &     & \checkmark        &          \\
			\cite{Chen_2023_Enhanced} & \checkmark       &     &    &       & \checkmark     &     &     &          &          \\
			\cite{Liu_2022_Item} & \checkmark       &     &    &       &      &     &\checkmark     &  \checkmark       &          \\
			\bottomrule
		\end{tabular}
	}
		\caption{Evaluation Criteria}	
\end{table}

Precision, Recall, and F1 are widely used to evaluate the accuracy of top-K recommendations. Precision@K measures the proportion of items clicked by the user among the recommended top-K items. Recall@K calculates the proportion of user clicks on the recommended top-K items compared to the entire set of clicks. F1@K is a combination of Precision@K and Recall@K.
\begin{equation}
	\begin{aligned}
		& \text { Precision@K }(u)=\frac{\left|R^K(u) \cap T(u)\right|}{K}, \quad \\
		& \text { Recall@K }(u)=\frac{\left|R^K(u) \cap T(u)\right|}{|T(u)|} \text {, } \\
		& \text { F1@K }(u)=\frac{2 \times \text { Precision } @ K(u) \times \text { Recall@ } K(u)}{\text { Precision } @ K(u)+\text { Recall@ } K(u)} . \\
		&
	\end{aligned}
\end{equation}

HR (Hit Rate) measures the proportion of users who have clicked on at least one recommended item, where T(u) represents the ground truth item set, $R^K(u)$ represents the top-K recommended item set, and I(·) is the indicator function.
\begin{equation}
	\text { HR@K }=\frac{1}{|\mathcal{U}|} \Sigma_{u \in \mathcal{U}} I\left(\left|R^K(u) \cap T(u)\right|>0\right)
\end{equation}

NDCG (Normalized Discounted Cumulative Gain) distinguishes the contribution of accurately recommended items based on their ranking position, where $(R_k^K(u)$ represents the k-th item in the recommendation list $R^K(u)$.
\begin{equation}
	\text { NDCG@K }=\frac{1}{|\mathcal{U}|} \sum_{u \in \mathcal{U}} \frac{\sum_{k=1}^K \frac{I\left(R_k^K(u) \in T(u)\right)}{\log (k+1)}}{\sum_{k=1}^K \frac{1}{\log (k+1)}}
\end{equation}

MAP (Mean Average Precision) is a widely adopted ranking metric that measures the average precision for users:

\begin{equation}
	\begin{aligned}
		\text{MAP@K} = & \frac{1}{|\mathcal{U}|} \sum_{u \in \mathcal{U}} \sum_{k=1}^K \frac{I(R_k^K(u) \in T(u)) \text{Precision@K}(u)}{K} \\
	\end{aligned}
\end{equation}

MRR (Mean Reciprocal Ranking) is an indicator used to evaluate the performance of ranking tasks, where the prediction with a higher rank corresponds to a larger reciprocal value, and a better score is reflected by the sum of these values:
\begin{equation}
	M R R=\frac{1}{|S|} \sum_{i=1}^{|S|} \frac{1}{\text { rank }_i}
\end{equation}

AUPRC (Area Under the Precision-Recall Curve) represents the area under the precision-recall curve, and its value ranges from 0 to 1. A value closer to 1 indicates better model performance.

TP represents true positive (the number of samples correctly predicted as positive by the model), FN represents false negative (the number of samples incorrectly predicted as negative by the model), FP represents false positive (the number of samples incorrectly predicted as positive by the model), and TN represents true negative (the number of samples correctly predicted as negative by the model).

Accuracy measures the overall prediction accuracy of the model on the entire sample set:
\begin{equation}
	\text { Accuracy }=\frac{T P+T N}{T P+T N+F P+F N}
\end{equation}

Hit Rate is another common performance evaluation metric for classification models, referring to the proportion of samples correctly predicted as positive by the model out of all samples:

\subsection{Comparison of the Results}

\begin{table}[]

	\scalebox{0.65}{ 	
		\begin{tabular}{cccc}
			\toprule
			Citation & Model Name & Year & Code Link                                                             \\\midrule
			\cite{Fatemi_2023_Learning}  & GISMo                                                            & 2023 & https://github.com/facebookresearch/gismo                       \\
			\cite{Ye_2023_Transformer-Based}  & -                                                                & 2023 & -                                                               \\
			\cite{Chen_2023_Enhanced} & EMRIGCN                                                          & 2023 & -                                                               \\
			\cite{Li__Food}  & -                                                                & 2022 & https://github.com/DiyaLI916/FoodKGE                            \\
			\cite{Lawrynowicz_2022_Food}  & -                                                                & 2022 & http://ontologydesignpatterns.org/wiki/Submissions:Food\_Recipe \\
			\cite{Zhang_2022_Learning}  & SCG-SPRe                                                         & 2022 & -                                                               \\
			\cite{Zhou_2022_Decoupled} & DHGAN                                                            & 2022 & https://github.com/wt-tju/DHGAN                                 \\
			\cite{Jian_2022_Multi-task}  & M-HetSage                                                        & 2022 & -                                                               \\
			\cite{Dai_2022_Decomposing}  & SPGCN                                                            & 2022 & https://github.com/lem0nle/SPGCN                                \\
			\cite{Liu_2022_Item} & IRGNN                                                            & 2022 & https://github.com/wwliu555/IRGNN\_TNNLS\_2021                  \\
			\cite{Yang_2022_Inferring}  & KAPR                                                             & 2021 & https://gitee.com/yangzijing\_flower/kapr/tree/master           \\
			\cite{Pellegrini_2021_Exploiting} & \begin{tabular}[c]{@{}c@{}}Food2Vec\\      FoodBERT\end{tabular} & 2021 & https://github.com/ChantalMP/                                   \\
			\cite{Shirai_2021_Identifying} & DIISH                                                            & 2020 & https://foodkg.github.io/subs.html                              \\
			\cite{Chen_2020_Try} & A2CF                                                             & 2020 & https://bit.ly/bitbucket-A2CF                                   \\
			\cite{Liu_2020_Decoupled} & DecGCN                                                           & 2020 & https://github.com/liuyiding1993/CIKM2020\_DecGCN               \\
			\cite{Chen_2020_Studying} & Product2Vec                                                      & 2020 & -                                                                                                                             \\
			\cite{Zhang_2019_Identifying}  & RRN                                                              & 2019 & -                                                               \\
			\cite{Shen_2019_KGDDS} & KGDDS                                                            & 2019 & http://www.iasokg.com/                                          \\
			\cite{Zhang_2019_Inferring} & SPEM                                                             & 2019 & -                                                               \\
			\cite{Rakesh_2019_Linked} & \begin{tabular}[c]{@{}c@{}}LVA\\      CLVA\end{tabular}          & 2019 & https://github.com/VRM1/WSDM19                                  \\
			\cite{Wang_2018_Path-constrained}  & PMSC                                                             & 2018 & -                                                               \\
			\cite{Ruiz_2019_SHOPPER} & SHOPPER                                                          & 2017 & https://github.com/franrruiz/shopper-src                        \\
			\cite{McAuley_2015_Inferring}  & Sceptre                                                          & 2015 & http://cseweb.ucsd.edu/$\sim$jmcauley/                          \\
			\cite{McAuley_2015_Image-Based}  & -                                                                & 2015 & http://cseweb.ucsd.edu/$\sim$jmcauley/.                         \\
			\cite{Boscarino_2014_Automatic}  & -                                                                & 2014 & -          \\\bottomrule                                                    
		\end{tabular}
	}
		\caption{Paper Code Link}
		\label{4_tab}
\end{table}

The table \ref{4_tab} below collects the code for the models compared in this review, along with a comparison of their results.

In most fields, datasets are widely distributed, and there is no universal dataset, making it difficult to make cross comparisons. In the e-commerce field, many models tend to choose Amazon's dataset for evaluating model performance. It is worth noting that Amazon's dataset also includes various types of goods. As table \ref{5_tab}, we present a comparison of the results of some models in terms of evaluation metrics such as Hit@10, NDCG@10, Accuracy, etc.

\begin{table*}
	\centering
	\begin{adjustbox}{angle=90, max width=\textwidth}
		\begin{tabular}{lllllll} \toprule
			Segments                        & Refs & NDGC@10 & HR@10  & Hit@10 & Accuracy & Precision \\ \midrule
			Women’s Clothing                & \cite{Wang_2018_Path-constrained}  & -       & -      & -      & -        & 0.9777    \\ 
			& \cite{McAuley_2015_Inferring}  & -       & -      & -      & 0.9587   & -         \\\midrule
			\multirow{2}{*}{Video Games}    & \cite{Zhang_2019_Inferring}  & -       & -      & 0.91   & -        & -         \\
			& \cite{Liu_2022_Item}  & -       & -      & -      & -        & 0.8177    \\\midrule
			Toys\_and\_Games                & \cite{Zhou_2022_Decoupled}  & 0.794   & 0.914  & -      & -        & -         \\\midrule
			Sports\_and\_Outdoors           & \cite{Zhou_2022_Decoupled}  & 0.745   & 0.866  & -      & -        & -         \\\midrule
			\multirow{2}{*}{Office}         & \cite{Chen_2020_Try}  & 0.254   & 0.3143 & -      & -        & -         \\
			& \cite{Wang_2018_Path-constrained}  & -       & -      & -      & -        & 0.9778    \\\midrule
			Musical Instrurnents            & \cite{Liu_2022_Item}  & -       & -      & -      & -        & 0.7977    \\\midrule
			\multirow{2}{*}{Men's Clothing} & \cite{Rakesh_2019_Linked}  & -       & -      & -      & 0.9282   & -         \\
			& \cite{McAuley_2015_Inferring}  & -       & -      & -      & 0.9669   & -         \\\midrule
			Home and Kitchen                & \cite{Wang_2018_Path-constrained}  & -       & -      & -      & -        & 0.8631    \\\midrule
			Grocery                         & \cite{Zhang_2022_Learning}  & 0.4097  & 0.6141 & -      & -        & -         \\\midrule
			\multirow{9}{*}{Electronics}    & \cite{Zhou_2022_Decoupled}  & 0.713   & 0.855  & -      & -        & -         \\
			& \cite{Yang_2022_Inferring}  & -       & -      & 0.87   & -        & -         \\
			& \cite{Liu_2020_Decoupled}  & 0.546   & 0.713  & -      & -        & -         \\
			& \cite{Zhang_2019_Identifying}  & -       & -      & -      & 0.9351   & -         \\
			& \cite{Zhang_2019_Inferring}  & -       & -      & 0.77   & -        & -         \\
			& \cite{Rakesh_2019_Linked}  & -       & -      & -      & 0.9547   & -         \\
			& \cite{Wang_2018_Path-constrained}  & -       & -      & -      & -        & 0.979     \\
			& \cite{McAuley_2015_Inferring}  & -       & -      & -      & 0.957    & -         \\
			& \cite{Liu_2022_Item}  & -       & -      & -      & -        & 0.757     \\\midrule
			\multirow{2}{*}{Clothing}       & \cite{Liu_2020_Decoupled}  & 0.472   & 0.631  & -      & -        & -         \\
			& \cite{Liu_2022_Item}  & -       & -      & -      & -        & 0.7428    \\\midrule
			\multirow{6}{*}{Cell Phone}     & \cite{Zhang_2022_Learning}  & 0.4542  & 0.6588 & -      & -        & -         \\
			& \cite{Yang_2022_Inferring}  & -       & -      & 0.78   & -        & -         \\
			& \cite{Chen_2020_Try}  & 0.198   & 0.3449 & -      & -        & -         \\
			& \cite{Zhang_2019_Identifying}  & -       & -      & -      & 0.9422   & -         \\
			& \cite{Zhang_2019_Inferring}  & -       & -      & 0.56   & -        & -         \\
			& \cite{Wang_2018_Path-constrained}  & -       & -      & -      & -        & 0.9016    \\\midrule
			\multirow{2}{*}{Books}          & \cite{Rakesh_2019_Linked}  & -       & -      & -      & 0.9571   & -         \\
			& \cite{McAuley_2015_Inferring}  & -       & -      & -      & 0.9376   & -         \\\midrule
			\multirow{4}{*}{Beauty}         & \cite{Yang_2022_Inferring}  & -       & -      & 0.94   & -        & -         \\
			& \cite{Liu_2020_Decoupled}  & 0.593   & 0.747  & -      & -        & -         \\
			& \cite{Zhang_2019_Identifying}  & -       & -      & -      & 0.8546   & -         \\
			& \cite{Zhang_2019_Inferring}  & -       & -      & 0.96   & -        & -         \\\midrule
			\multirow{5}{*}{Baby}           & \cite{McAuley_2015_Inferring}  & -       & -      & -      & 0.9218   & -         \\
			& \cite{Zhang_2022_Learning}  & 0.292   & 0.4857 & -      & -        & -         \\
			& \cite{Yang_2022_Inferring}  & -       & -      & 0.89   & -        & -         \\
			& \cite{Zhang_2019_Identifying}  & -       & -      & -      & 0.8975   & -         \\
			& \cite{Zhang_2019_Inferring}  & -       & -      & 0.89   & -        & -         \\\midrule
			Automovie                       & \cite{Chen_2020_Try}  & 0.1788  & 0.2978 & -      & -        & -        \\\bottomrule  
		\end{tabular}
	\end{adjustbox}
	\caption{Amazon dataset results comparison}
		\label{5_tab}
\end{table*}

\section{FUTURE DIRECTIONS AND PROSPECTS}
Delving deeper into the study , the prospects for enhancing substitute recommendation systems appear promising. Through ongoing research and innovation, there is an opportunity to significantly improve the personalization and accuracy of substitute recommendations across various domains. This paves the way for the continued development and widespread application of substitute relationship analysis, opening up new avenues for exploration and advancement in this field.

\subsection{Complex Substitute Relationship Modeling}
\subsubsection{Substitute relationship modeling in field of commodities}
\textbf{Fashion Recommendation} considers both the similarity and takes compatibility into account  \cite{Gu2022PAINTPF,10.1145/3591106.3592224,Mohammadi2021SmartFA}. This is a challenging task because it often requires the use of information from different sources, such as shaping fashion influences from photographs \cite{AlHalah2020ModelingFI}.   The future challenge lies in how to introduce more realistic try-on effects, such as using 3D virtual fitting technology to allow users to better preview the effect of products.

In the case of \textbf{electrical appliances}, substitute relationships mainly involve matching at the functional level. A future challenge is how to infer from the functional characteristics of electrical products to help users find substitute products. Another challenge in substitute relationship inference is effectively balancing the relationships among product functionality, performance, and price. This means that it is necessary to develop models that can consider multiple factors comprehensively .

Artistic works such as books, movies, and music can be collectively referred to as cultural and entertainment works. In the context of \textbf{substitute recommendations for cultural and entertainment works}, the characteristics of the content need to be considered.
\begin{itemize}
	\item[$\bullet$]Book recommendation: In book recommendation, it is necessary to consider semantic-level data features. How to accurately capture users' subjective preferences and personalized needs is a challenge in building personalized book recommendation systems.
	\item[$\bullet$]Movie recommendation: In movie recommendation, in addition to considering semantic features such as plot and theme, factors such as visual and audio styles can also be taken into account, by using movie tags, keywords, or metadata. 
	\item[$\bullet$]Music recommendation: The style, melody, rhythm, and instrumental performance of music are all important factors that determine the similarity between pieces of music. Music feature extraction and audio analysis techniques can be used to calculate the similarity of music.
\end{itemize}

\subsubsection{Other fields}

Modeling cross-domain substitution relationships faces unique challenges and opportunities\cite{9055059, Wu2022EAGCNAE}. When conducting substitute recommendations in non-e-commerce fields such as academia, tourism, and healthcare, specific problems and challenges are encountered.

\begin{itemize}
	\item[$\bullet$]\textbf{Academic literature} covering multiple disciplinary fields , often characterized by high specialization and complexity. Therefore, when conducting substitute recommendations, it is necessary to consider how to accurately understand and match the professional requirements of different disciplinary fields. 
	\item[$\bullet$]In the \textbf{tourism field}, it may involve text-based destination introductions, travel guides. Substitute recommendations need to take into account seasonal factors , allowing users to choose the right destination at the right time. This can involve introducing cross-regional samples during the training process or using data synthesis techniques to generate samples with regional characteristics \cite{Hacheme2021NeuralFI}. 
	\item[$\bullet$]Image-based substitute recommendations may involve \textbf{medical image diagnostics}, necessitating the assurance that the recommended image data meets the requirements for diagnostic accuracy and timeliness in clinical applications.
\end{itemize}

\textbf{Recommendations in the field of diet} need to be based on reliable and accurate data sources, such as nutritional components and calorie information of ingredients. It is necessary to comprehensively consider factors such as nutritional balance, taste, health, and satiety. This involves the problem of multi-objective optimization and trade-offs. Balancing different needs is a difficult task. 

\subsubsection{Multimodal and Multilevel Relationship Modeling}
By considering both multimodal and multilevel relationship modeling, we can more comprehensively and accurately represent substitute relationships between entities.

In \textbf{multimodal relationship modeling}, various data types can be combined to model substitute relationships. For example, semantic information from textual data can be fused with visual features from image data \cite{Wu2020ModelingPV,Hsiao2021FromCT} . Cross-modal fusion enables the interaction of information between different data types, providing richer feature and semantic expressions in relationship modeling \cite{Zhang2021GeometrySC}.

\textbf{Multilevel relationship modeling} considers the various levels and complexities of relationships between entities \cite{10.1145/3159652.3159680}. This can be achieved by establishing a multilevel relationship network, where each level corresponds to different relationship types or levels of relationship abstraction. For example, in a knowledge graph, relationships can be categorized into different hierarchical levels such as parent-child relationships, attribute relationships, and so on, and graph neural networks or multilevel attention mechanisms can be used to model these relationships \cite{Patil2021AGT}. In this way, we can better capture relationships at different levels, leading to more accurate substitute relationship reasoning.

\textbf{Combining substitutions} can lead to more accurate and practical results.One-to-one substitutions may not always meet the users' needs. For example, when a user wants to substitute chocolate, they may require cocoa powder, sugar, and butter. Combination substitutions also faces some challenges, encounter challenges such as compatibility, coordination, manufacturing costs, and user acceptance. Items need adjustments to work well together and might raise manufacturing complexity.

\subsection{Considering Temporality and Dynamics}

\subsubsection{User Intent Extraction}
Context-based substitute reasoning refers to inferring the current needs and intentions of users based on their context and background information, such as geographic location, time, personal preferences, etc., and then making substitute recommendations accordingly. The approach to extracting user intent involves using deep learning-based large language models and intent recognition, as well as mining algorithms based on user behavior and interests. By analyzing users' browsing history, purchase records, or rating data \cite{Doniec2020PurchaseIA,Ding2022ModelingIU,He202BARBR}. At the same time, establishing real-time interaction with users is also a key.

When there are shifts or \textbf{multiple complex intents in a user's dialogue}, this is because in such cases, the user's expressions may be ambiguous or unclear, with multiple intents intertwined or overlapping.  By incorporating user feedback, continuously optimizing intent recognition models, and utilizing contextual information, the system can achieve more accurate intent classification.

\subsubsection{The problem of sequence recommendation}
In the real world, there are sequential dependencies between many events and objects. For example, sequence recommendation can help users adapt to temporal changes and is widely used in fields like e-commerce, news, and social media\cite{Sun2022SequentialGC, Zhang2022GatingAC, Wu2022GCRecGC, Ding2022ModelingIU, Pang2021ModelingFC}. User interests can include long-term and short-term preferences. Session-based recommendation systems\cite{Hidasi_2018_Recurrent, Wu_2019_Session-Based, Meng_2020_Incorporating, Wang_2022_Effectively} capture users' short-term preferences from their recent sessions and reflect preference dynamics from one session to another, thereby providing more accurate and timely recommendations.

However, the current relationship modeling models face interpretability issues. To address this, researchers have been working on developing model interpretation techniques and designing models with strong interpretability\cite{10.1145/3404835.3462911, 10.1145/3437963.3441754, Rago2021ArgumentativeEF}. Model interpretation techniques include rule-based explanations, local sensitivity analysis, attention mechanisms, etc., which can help reveal the basis and reasoning process behind model decisions. In addition, designing models with strong interpretability is also important. For example, using graph neural networks to model relationships in graph data can provide more intuitive explanations through node and edge representations.

\subsection{Integration with Large Language Models}

Large language models, such as GPT (Generative Pre-trained Transformer), have become a popular research direction. These models have a large number of parameters and are pre-trained on massive amounts of data, making them applicable to various tasks such as question answering, sentiment analysis, text classification, and more.

Large language models offer several advantages in relationship modeling. They can learn language patterns, capture rich semantic information, and have broad applications in substituting relational reasoning. However, these models also have limitations, with the most important one being the need for substantial amounts of training data. Additionally, the performance of large language models on specific tasks can be influenced by issues like data skew and domain differences, requiring further optimization and adaptation to achieve better results. Therefore, when using large language models for relationship modeling, it is important to consider both their strengths and limitations and thoroughly evaluate their suitability for the specific task at hand.

\subsubsection{Data-driven Relationship Reasoning}
Large language models \textbf{exhibit biases in their training data}, leading to the generation of results that may contain discriminatory, biased, or unfair content. In many fields, \textbf{obtaining large-scale annotated data} is a time-consuming and costly task. The advantage of unsupervised learning is that it can learn from unlabeled data and discover patterns without relying on a large amount of labeled data \cite{Wada2022UnsupervisedLS}. Unsupervised learning faces challenges such as distribution shift of data and label scarcity. To overcome these challenges some potential solutions can be proposed \cite{Chen2020CoSamAE,Chaudhuri2021SemisupervisedSO}. \textbf{The ``few-shot" problem} refers to a situation in machine learning and pattern recognition where a class has a limited number of samples, making it challenging for the model to grasp the characteristics and patterns of that class sufficiently during the learning process \cite{10.1145/3477495.3531978,10.1145/3580305.3599515,10.1145/3583780.3615110}. Substitute reasoning can address the few-shot problem by generating new samples through analogy reasoning or inference between samples, thereby expanding the quantity and diversity of existing data.

\subsubsection{Generative Reasoning}

Large language models are typically based on statistical modeling and lack true reasoning abilities. Generative reasoning is an inference approach based on logic and reasoning rules, aimed at deriving logically correct conclusions that are more flexible, rigorous, and expressive. Integrating generative reasoning with large language models allows for the combination of existing knowledge graphs or knowledge bases with the models. By incorporating external knowledge, reasoning can be supported, thereby enhancing the accuracy and logic of generative reasoning. When combined with substitute reasoning, for example, when recommending a product to a user, the model can generate a text that describes the characteristics and advantages of the product, explaining why it is a good choice for the user.

\section{CONCLUSIONS}
Substitution relationships play a crucial role in various aspects of daily life, from cooking to consumer demands. The ability to identify appropriate substitutions is a valuable skill that individuals must possess, as it enables them to adapt and overcome challenges while maintaining the desired outcomes. This reasoning for substitution relationships is essential in both personal and professional settings, as it aids in decision-making and problem-solving.

To delve deeper into substitution relationships, research methods have been developed to understand and predict these associations. Machine learning algorithms, natural language processing techniques, and semantic analysis tools have all contributed to enhancing our understanding of substitution reasoning. These methods analyze large datasets, including recipes, customer feedback, and historical buying patterns, to uncover patterns and establish substitution relationships. By combining qualitative and quantitative approaches, researchers can gain insights into the factors that influence substitution choices and develop accurate predictive models.

In conclusion, reasoning for substitution relationships is a fundamental aspect of daily life, influencing various domains such as cooking, consumer demands, and recommendation systems. It empowers individuals to adapt and make informed choices, while also enabling businesses to meet customer needs. Through research and innovation, we can continue to enhance personalization and accuracy in substitution recommendations, further improving efficiency and customer satisfaction in industries like e-commerce and food. Substitution reasoning is a dynamic field with vast potential, and further exploration and development in this area will undoubtedly lead to exciting advancements in the future.

\vskip 2mm
\zihao{5}
\noindent
\textbf{Acknowledgment}
\vskip 2mm

\zihao{5--}
\noindent
	This review study was carried out with the support of the National Science Foundation of China (No.62162048,62262047), the Self-topic Project of Engineering Research Center of Ecological Big Data, the Natural Science Foundation of Inner Mongolia in China (No.2021MS01023), the National Natural Science Foundation of China (No.62062052), Ministry of Education, and Inner Mongolia Science and Technology Plan Project (2021GG0164).

\vskip 2mm
\zihao{5}
\noindent

\bibliographystyle{plain}
\bibliography{yax}

\clearpage
\clearpage
\begin{strip}

\end{strip}

\begin{biography}[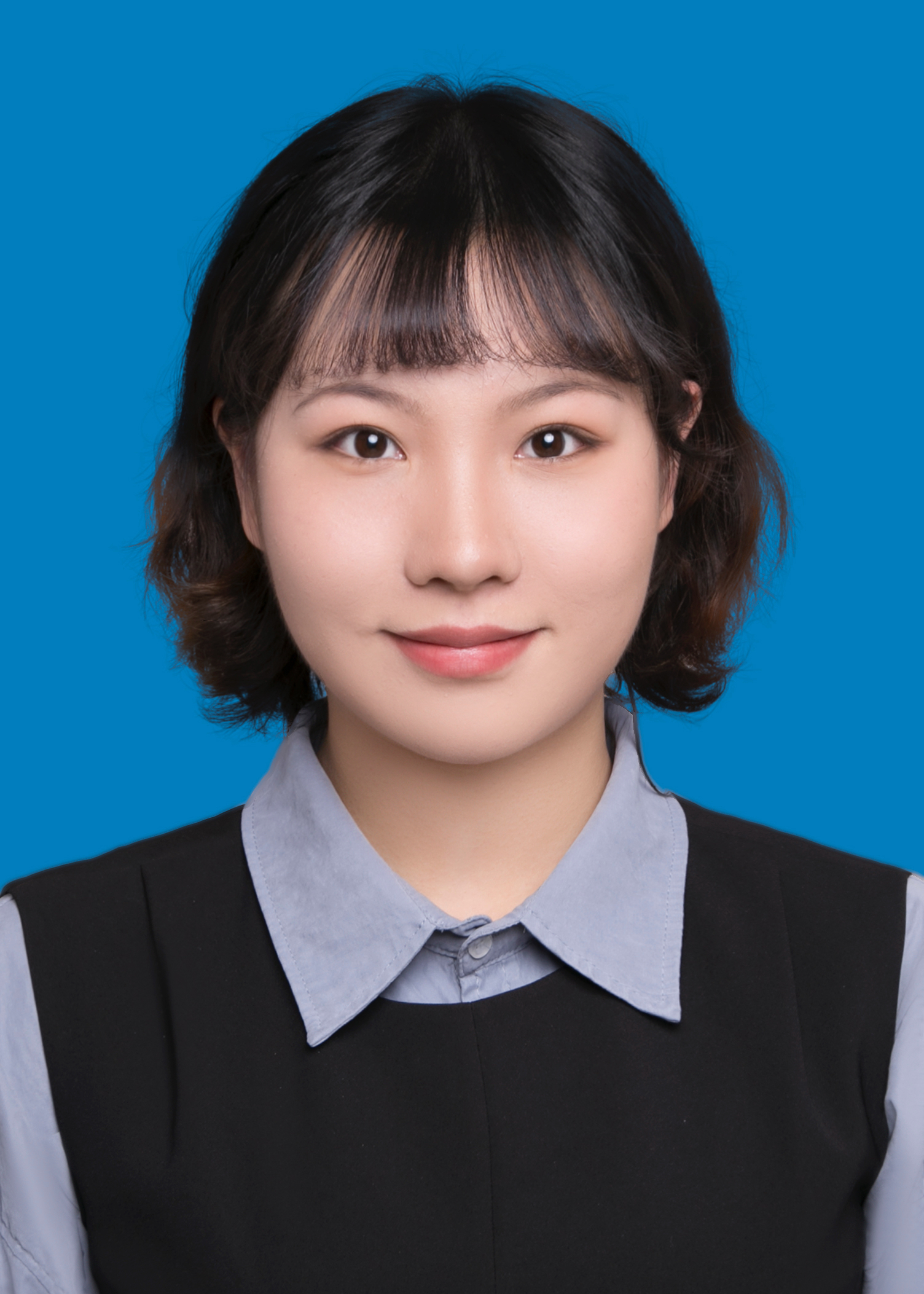]
	\noindent
	\textbf{Anxin Yang} obtained a Bachelor’s degree in Software Engineering from Inner Mongolia University. She is currently a graduate student at the School of Computer Science, Inner Mongolia University. Her main research areas are data mining and recommendation systems.
\end{biography}

\begin{biography}[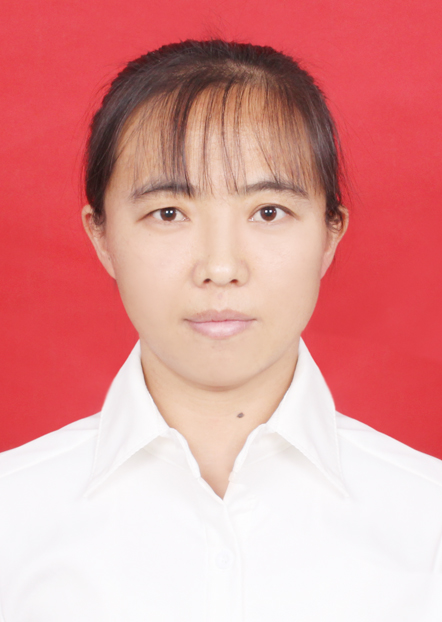]
	\noindent
	\textbf{Zhijuan Du} obtained her Ph.D. degree from Renmin University of China in 2018. She currently holds the position of associate professor and master’s supervisor. She focuses on researching knowledge graphs, intelligent assignment, recommendation systems, deep learning, and social big data mining.
\end{biography}

\begin{biography}[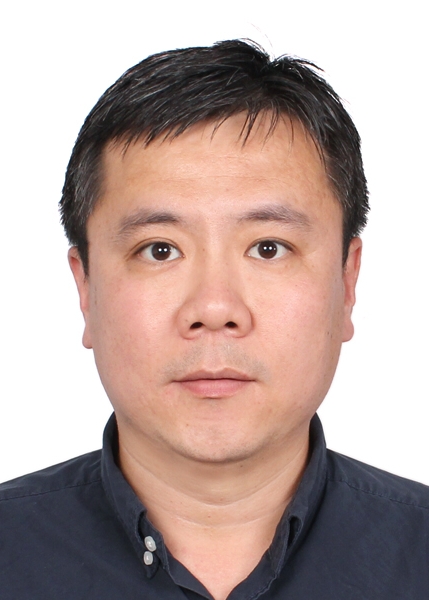]
	\noindent
	\textbf{Tao Sun} obtained a Ph.D. in Computer Science from the School of Computer Science, Inner Mongolia University in 2013, and currently serves as a professor at the College of Computer Science. He focuses on researching formal methods and software testing.
\end{biography}

  \end{document}